\documentclass[usenatbib]{mnras}

\usepackage{newtxtext,newtxmath}

\usepackage[T1]{fontenc}
\usepackage{ae,aecompl}

\usepackage{graphicx}	
\usepackage{amsmath}	
\usepackage{amssymb}	
\usepackage[usenames, dvipsnames]{color}
\usepackage{url}
\usepackage{booktabs}

\title[Massive Galaxies at Cosmic Noon]{Exploring the High-Mass End of the Stellar Mass Function of Star Forming Galaxies at Cosmic Noon}

\author[Sherman et al.]{Sydney Sherman$^{1}$\thanks{E-mail: \texttt{ssherman@astro.as.utexas.edu}},
Shardha Jogee$^{1}$,
Jonathan Florez$^{1}$,
Matthew L. Stevans$^{1}$, \newauthor
Lalitwadee Kawinwanichakij$^{2,3,4}$, 
Isak Wold$^{1,5}$,
Steven L. Finkelstein$^{1}$, 
Casey Papovich$^{2,3}$, \newauthor
Viviana Acquaviva$^{6}$,
Robin Ciardullo$^{7,8}$, 
Caryl Gronwall$^{7,8}$, 
and Zacharias Escalante$^{1}$
\\
\\
$^{1}$Department of Astronomy, The University of Texas at Austin, Austin, TX 78712 \\
$^{2}$Department of Physics and Astronomy, Texas A\&M University, College Station, TX, 77843-4242\\
$^{3}$George P. and Cynthia Woods Mitchell Institute for Fundamental Physics and Astronomy, Texas A\&M University, College Station, TX 77843\\
$^{4}$LSSTC Data Science Fellow\\
$^{5}$NASA Goddard Space Flight Center, Greenbelt, MD 20771\\
$^{6}$Department of Physics, New York City College of Technology, Brooklyn, NY 11201\\
$^{7}$Department of Astronomy and Astrophysics, The Pennsylvania State University, University Park, PA 16802\\
$^{8}$Institute for Gravitation and the Cosmos, The Pennsylvania State University, University Park, PA 16802
}

\date{Accepted 2019 November 8. Received 2019 November 8; in original form 2019 February 22}

\pubyear{2019}

\begin{document}
\label{firstpage}
\pagerange{\pageref{firstpage}--\pageref{lastpage}}
\maketitle

\begin{abstract}
We present the high-mass end of the galaxy stellar mass function using the largest sample to date (5,352) of star-forming galaxies with $M_{\star} > 10^{11} M_{\odot}$ at cosmic noon, $1.5 < z < 3.5$. This sample is uniformly selected across 17.2 deg$^2$ ($\sim$0.44 Gpc$^3$ comoving volume from $1.5 < z < 3.5$), mitigating the effects of cosmic variance and encompassing a wide range of environments. This area, a factor of 10 larger than previous studies, provides robust statistics at the high-mass end. Using multi-wavelength data in the \textit{Spitzer}/HETDEX Exploratory Large Area (SHELA) footprint we find that the SHELA footprint star-forming galaxy stellar mass function is steeply declining at the high-mass end probing values as high as $\sim$$10^{-4}$ Mpc$^3$/dex and as low as $\sim$5$\times$$10^{-8}$ Mpc$^3$/dex across a stellar mass range of log($M_\star$/$M_\odot$) $\sim$ 11 - 12. We compare our empirical star-forming galaxy stellar mass function at the high mass end to three types of numerical models: hydrodynamical models from IllustrisTNG, abundance matching from the UniverseMachine, and three different semi-analytic models (SAMs; SAG, SAGE, GALACTICUS). At redshifts $1.5 < z < 3.5$ we find that results from IllustrisTNG and abundance matching models agree within a factor of $\sim$2 to 10, however the three SAMs strongly underestimate (up to a factor of 1,000) the number density of massive galaxies. We discuss the implications of these results for our understanding of galaxy evolution.
\end{abstract}



\section{Introduction}
Massive galaxies present an excellent testbed of galaxy evolution and structure formation theories. Evidence suggests that massive galaxies are home to old stellar populations (\citealt{Cowie1996}, \citealt{Fontanot2009}, \citealt{Fontana2009}, \citealt{Kajisawa2011}, \citealt{Greene2015}, among others) that formed rapidly at early times, developing first as compact red nuggets (e.g., \citealt{vanderWel2011}, \citealt{Weinzirl2011}, \citealt{vanDokkum2015}) and building extended populations at later times through minor mergers, which are known to be frequent at $z < 2$ (e.g., \citealt{Jogee2009}, \citealt{Lotz2011}). Despite their importance, massive galaxies have proven challenging to study at high redshifts due to small-area surveys (e.g., \citealt{Grogin2011}, \citealt{Koekemoer2011}, \citealt{Whitaker2011}, \citealt{Straatman2016}) that find only a handful of these elusive objects. As a result, the assembly history of the most massive galaxies remains largely unclear. 

Measures of the number of galaxies at different epochs, typically in the form of the galaxy stellar mass function (\citealt{Muzzin2013}, \citealt{Ilbert2013}, \citealt{Tomczak2014}, \citealt{Grazian2015}, \citealt{Davidzon2017}, among others), have become benchmark tests for galaxy formation theory (e.g., \citealt{Behroozi2010}, \citealt{Vogelsberger2014a}, \citealt{Somerville2015}, \citealt{Pillepich2018b}). Assembling galaxies at the high-mass end, in particular, requires efficient formation mechanisms such as major mergers and high star formation rates (SFRs) to build such massive systems by $z \sim 2$. Therefore, the number density of high-mass galaxies places constraints on the merger history and star-formation efficiency of these systems, and can strongly inform theoretical models of galaxy evolution. 

On the theoretical front, different classes of numerical models, such as hydrodynamical simulations, semi-analytic models (SAMs), and abundance matching models are making increasingly sophisticated attempts to simulate massive galaxies from early cosmic times to today. Hydrodynamical simulations, such as IllustrisTNG (\citealt{Pillepich2018b}, \citealt{Springel2018}, \citealt{Nelson2018}, \citealt{Naiman2018}, \citealt{Marinacci2018}) are now covering larger volumes (IllustrisTNG simulation suite includes $\sim$50$^3$ Mpc$^3$ to $\sim$300$^3$ Mpc$^3$ volumes) and including more sophisticated implementations of feedback processes \citep{Weinberger2017}. \textit{Multi-DarkPlanck2} and \textit{Bolshoi-Planck} dark-matter-only simulations (e.g., \citealt{Klypin2016}, \citealt{RodriguezPuebla2016}, \citealt{Knebe2018}) simulate much larger volumes of 0.25$h^{-1}$Gpc and 1.0 $h^{-1}$Gpc, respectively, on a side with different resolutions. The properties they output for the dark-matter component can be combined with different implementations of baryonic physics in SAMs (e.g., \citealt{Benson2012}, \citealt{Somerville2015}, \citealt{Croton2016}, \citealt{Cora2018}) or in abundance matching models (e.g., \citealt{Behroozi2010}, \citealt{Behroozi2019}) to model galaxy evolution. 

Previous high redshift studies have focused on very deep, small-area surveys (e.g., \citealt{Grogin2011}, \citealt{Koekemoer2011}, \citealt{Whitaker2011}, \citealt{Straatman2016}) pushing constraints on the stellar mass function to the low mass end. Deep, small-area surveys, however, suffer from the effects of cosmic variance at the high-mass end (e.g., \citealt{Driver2010}, \citealt{Moster2011}) and find only small samples of high-redshift galaxies with $M_{\star} > 10^{11} M_{\odot}$ (e.g., \citealt{Conselice2011}, \citealt{Weinzirl2011}, \citealt{Wang2012}, \citealt{Muzzin2013}, \citealt{Ilbert2013}, \citealt{Tomczak2014}). Works such as \cite{Muzzin2013}, \cite{Ilbert2013}, and \cite{Tomczak2014} closely examine evolution in the stellar mass function, finding that the quiescent population evolves rapidly since $z \sim 2$, while there is little evolution in the star-forming population. In the local Universe, recent integral field spectroscopy (IFS) studies (e.g., \citealt{Sanchez2012}, \citealt{Ma2014}, \citealt{Bundy2015}) have confirmed that the most massive galaxies in the nearby Universe have old stellar populations indicative of early growth and rapid quenching at early times (e.g., \citealt{Perez2013}, \citealt{Greene2015}). Deep, wide-area surveys (e.g., \citealt{Jannuzi1999}, \citealt{deJong2015}, \citealt{Papovich2016}, \citealt{Wold2019}) are necessary to compile large, statistically significant samples of rare systems, such as massive galaxies, overcome cosmic variance, and probe a wide range of environments. As massive galaxies present a challenge to the overall hierarchical structure formation expected in $\Lambda$CDM cosmology (e.g., \citealt{Cowie1996}, \citealt{Fontanot2009}), it is imperative that statistically significant samples of massive galaxies are constructed. 

Here, we present an unprecedentedly large sample of massive star-forming galaxies (5,352 galaxies with $M_{\star} > 10^{11} M_{\odot}$) in the \textit{Spitzer}/HETDEX Exploratory Large Area (SHELA) footprint, a $\sim$17.2 deg$^2$ region of SDSS Stripe 82 with extensive multi-wavelength coverage from the rest-frame UV through far-IR/submillimeter. Our study focuses on the peak in cosmic star formation rate density ($1.5 < z < 3.5$), a time when star formation and black hole accretion peaked, proto-clusters began to collapse, and galaxies underwent significant growth. With such a large area survey and a uniformly selected sample, we are able to place some of the most robust constraints to date on the high-mass end of the star-forming galaxy stellar mass function with negligible uncertainties due to cosmic variance. 

This paper is organized as follows. In Section \ref{sec:data} we introduce the data available in the SHELA footprint and the data products used in this work. The analysis used to obtain photometric redshifts, galaxy properties, and selection of a galaxy sample is discussed in Section \ref{sec:analysis}. We introduce the SHELA footprint star-forming galaxy stellar mass function in Section \ref{sec:smf} where we compare this stellar mass function to previous results from observation and theory and discuss the implications of measuring the stellar mass function in large area surveys. Finally, we summarize our results in Section \ref{sec:summary}. Throughout this work we adopt a flat $\Lambda$CDM cosmology with $h = 0.7$, $\Omega_m = 0.3$, and $\Omega_{\Lambda} = 0.7$. 

\section{Data}
\label{sec:data}
This work utilizes a unique large-area, multi-wavelength survey conducted in the data-rich equatorial field SDSS Stripe 82 to study the significant epoch of $1.5 < z < 3.5$. The SHELA/HETDEX survey footprint (\citealt{Papovich2016}, \citealt{Wold2019}), consists of five photometric data sets (Dark Energy Camera (DECam) \textit{u,g,r,i,z} \citep{Wold2019}, NEWFIRM $K_s$ (PI Finkelstein; Stevans et al.\ in preparation), \textit{Spitzer}-IRAC 3.6 and 4.5$\mu$m (PI Papovich; \citealt{Papovich2016}), \textit{Herschel}-SPIRE (HerS, \citealt{Viero2014}) far-IR/submillimeter, as well as XMM-Newton and Chandra X-ray Observatory X-ray data from the Stripe 82X survey (\citealt{LaMassa2013a}, \citealt{LaMassa2013b}, \citealt{Ananna2017})) and a future optical spectroscopic survey from the Hobby Eberly Telescope Dark Energy Experiment (HETDEX, \citealt{Hill2008}). Additional $J$ and $K_s$ data from the VICS82 Survey \citep{Geach2017} are used to supplement the near-IR data available in the SHELA region. 

In this work, we utilize the DECam \textit{u,g,r,i,z}, VICS82 $J$ and $K_s$, and \textit{Spitzer}-IRAC 3.6 and 4.5$\mu$m data in the SHELA footprint to obtain photometric redshifts and galaxy properties, such as stellar mass and star formation rate, for $1.5 < z < 3.5$ galaxies. This redshift range probes the peak of the cosmic star formation rate density \citep{MadauDickinson2014}, and also contains the redshift range ($1.9 < z < 3.5$) over which the HETDEX project is expected to obtain Ly$\alpha$-based spectroscopic redshifts for approximately one third of our sources \citep{Hill2008}. 
This rest-frame UV to rest-frame near-IR subset of the full data products within the SHELA footprint covers $\sim$17.2 deg$^2$, corresponding to $\sim$0.44 Gpc$^3$ from $1.5 < z < 3.5$. This huge comoving volume mitigates the effects of cosmic variance and allows for the compilation of a statistically significant sample of massive galaxies (5,352 star-forming galaxies with $M_{\star} > 10^{11} M_{\odot}$) across diverse environments (e.g., group-like, proto-cluster, field). 

A complete description of the data reduction procedure for both the DECam and \textit{Spitzer}-IRAC data can be found in \cite{Wold2019}, and will very briefly be described here. The DECam \textit{u,g,r,i,z} catalog is built from a stacked, PSF-matched $r,i,z$ detection image using Source Extractor (SExtractor; \citealt{BertinArnouts1996}). By using $r,i,z$ selection, our sample is biased towards star-forming galaxies, and we describe how any remaining quiescent galaxies are removed from our sample in Section \ref{sec_highConfidence}. In this work, we utilize the MAG\_AUTO values output by SExtractor. The DECam data reach a $5\sigma$ depth of r = 24.5 AB mag \citep{Wold2019}. IRAC photometry is extracted using The Tractor \citep{tractor}, which uses forced photometry to deblend the IRAC imaging with source positions from the higher resolution DECam catalog used as location priors. This procedure is necessary due to the low spatial resolution of IRAC ($\sim$1.9$\arcsec$) compared to that of DECam ($\sim$1.2$\arcsec$) and improves upon the default SExtractor deblending procedure that was used in the original SHELA \textit{Spitzer}-IRAC catalog from \cite{Papovich2016}. The IRAC data reach a $5\sigma$ depth of 22 AB mag in both IRAC bands. Additionally, we use SExtractor MAG\_AUTO values in the $J$ and $K_s$ bands from the VICS82 Survey, which reach $5\sigma$ depths of 21.4 and 20.9 AB mag respectively \citep{Geach2017}. 

\begin{figure}
\begin{center}
\includegraphics[width=\columnwidth]{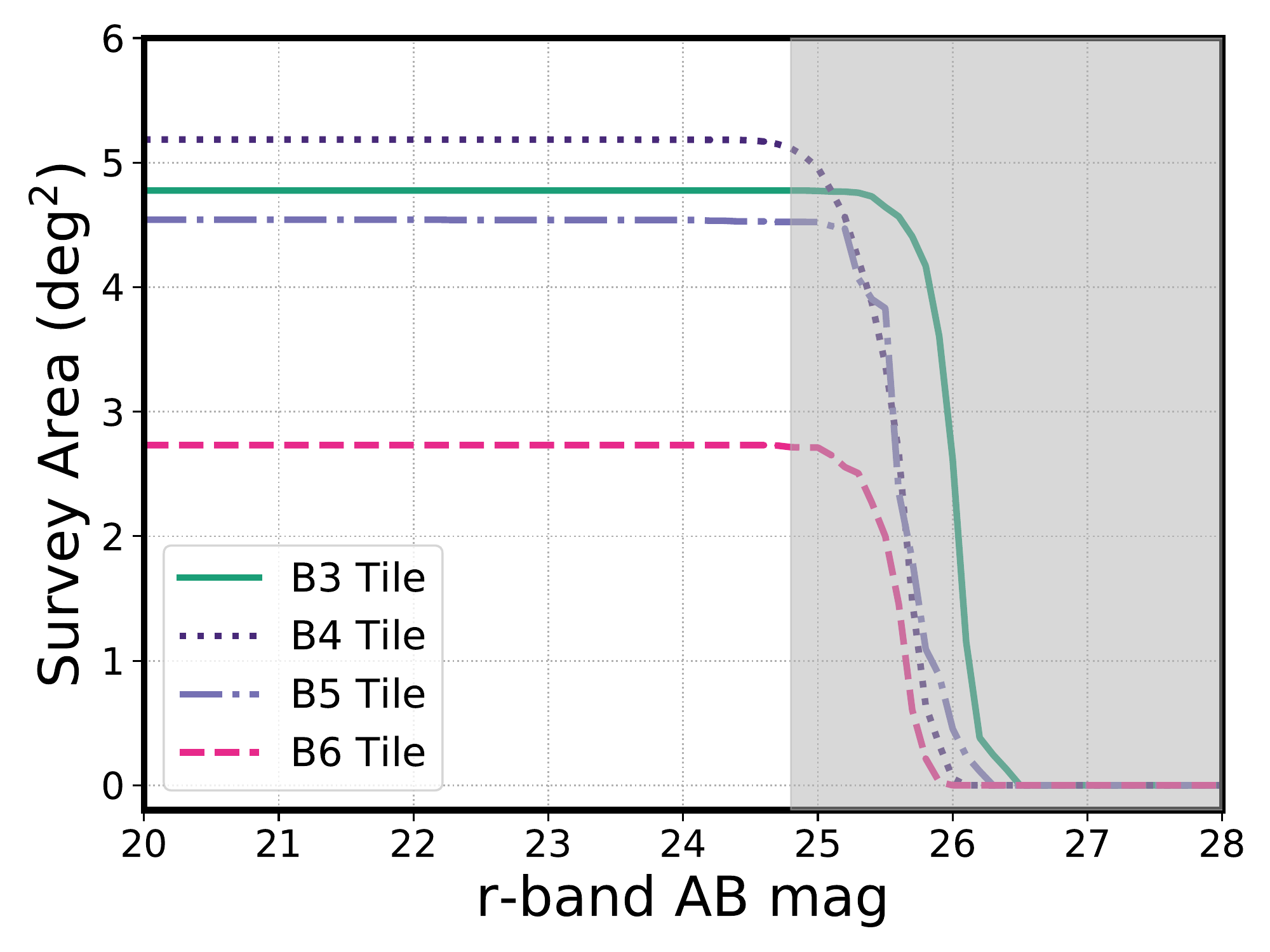}
\caption{The area in which an object with a given $r$-band magnitude can be detected in our survey footprint with S/N $\geq$ 5. The gray shaded region represents magnitudes fainter than the $r$ = 24.8 AB mag 80\% completeness limit found by \protect\cite{Wold2019}. Across all four adjoining tiles (B3, B4, B5, and B6) that comprise the SHELA footprint, the area does not vary as a function of magnitude for objects brighter than the $r$-band 80\% completeness limit. Across all four tiles, objects can be detected with $r$-band S/N $\geq$ 5 in $\sim$17.2 deg$^2$.  }
\label{survey_area}
\end{center}
\end{figure}

To determine the area of our survey footprint, we count the number of pixels in the DECam $r$-band image where an object with a particular $r$-band magnitude could be found with S/N $\geq$ 5. This is done using the $r$-band to conform to the $r,i,z$-based selection for the DECam catalog. The DECam imaging \citep{Wold2019} is divided into four adjoining tiles denoted B3, B4, B5, and B6, and we compute the area separately for each tile. For $r$-band magnitudes spanning $r$ = 20 - 28 AB mag with step size 0.1 mag, we determine the threshold noise in nJy required for a source with that magnitude to have S/N = 5. If a given pixel in the DECam rms map (containing values for the background noise associated with each pixel) has 5$\sigma$ rms value larger than the S/N = 5 threshold for an object with a given $r$-band magnitude, then that pixel is not included in the calculated survey area for objects of that magnitude. We find that the survey area within each tile (Figure \ref{survey_area}) does not evolve as a function of magnitude for all objects brighter than the $r$ = 24.8 AB mag 80\% completeness limit reported by \cite{Wold2019}. With all four tiles combined, objects above the $r$-band 80\% completeness limit can be detected with S/N $\geq$ 5 across $\sim$17.2 deg$^2$. 

\section{Data Analysis And SED Fitting}
\label{sec:analysis}
\subsection{Galaxy Properties From EAZY-py}
\subsubsection{Photometric Redshifts and Error Estimates}
\label{sec:photo-z}
\begin{figure}
\begin{center}
\includegraphics[width=\columnwidth]{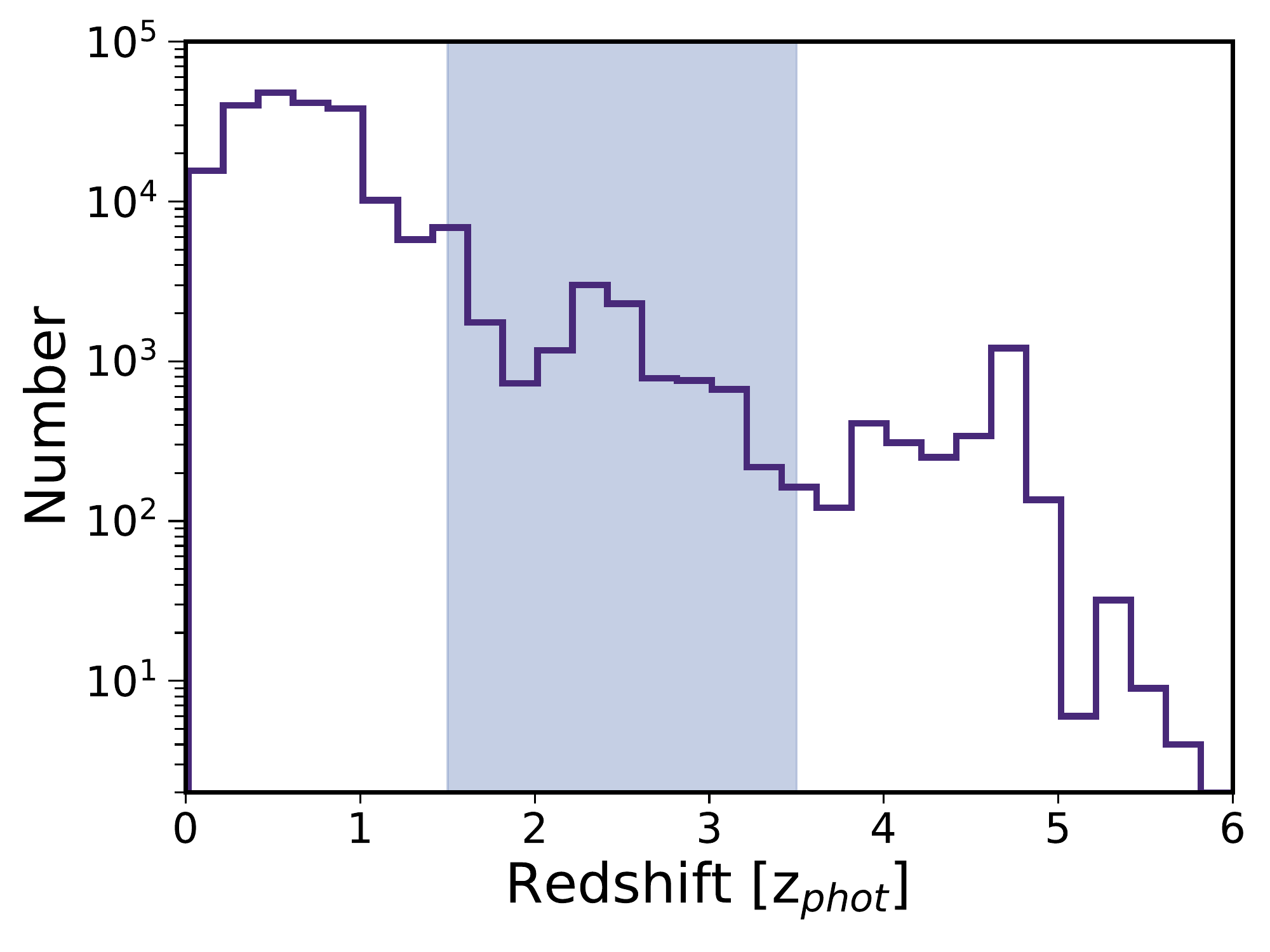}
\caption{Photometric redshift distribution for high-confidence star-forming galaxies (see Section \ref{sec_highConfidence} for a description of sample selection) in the SHELA footprint. The photometric redshift shown here, and used throughout this paper, refers to the EAZY-py best-fit redshift at which $\chi^2$ is minimized. The shaded region represents the redshift range of interest ($1.5 < z < 3.5$) in this paper. For z < 1 galaxies with spectroscopic redshifts from SDSS, we find $\sigma_{\rm NMAD}$ = 0.0377. We discuss tests of photometric redshift recovery of $1.5 < z < 3.5$ galaxies using a sample of mock galaxies in Section \ref{sec:fake_galaxies}. Across all redshifts, we find 219,996 galaxies, of which, 14,910 have redshifts $1.5 < z < 3.5$. }
\label{zphot_hist}
\end{center}
\end{figure}

\begin{figure*}
\begin{center}
\includegraphics[width=\textwidth]{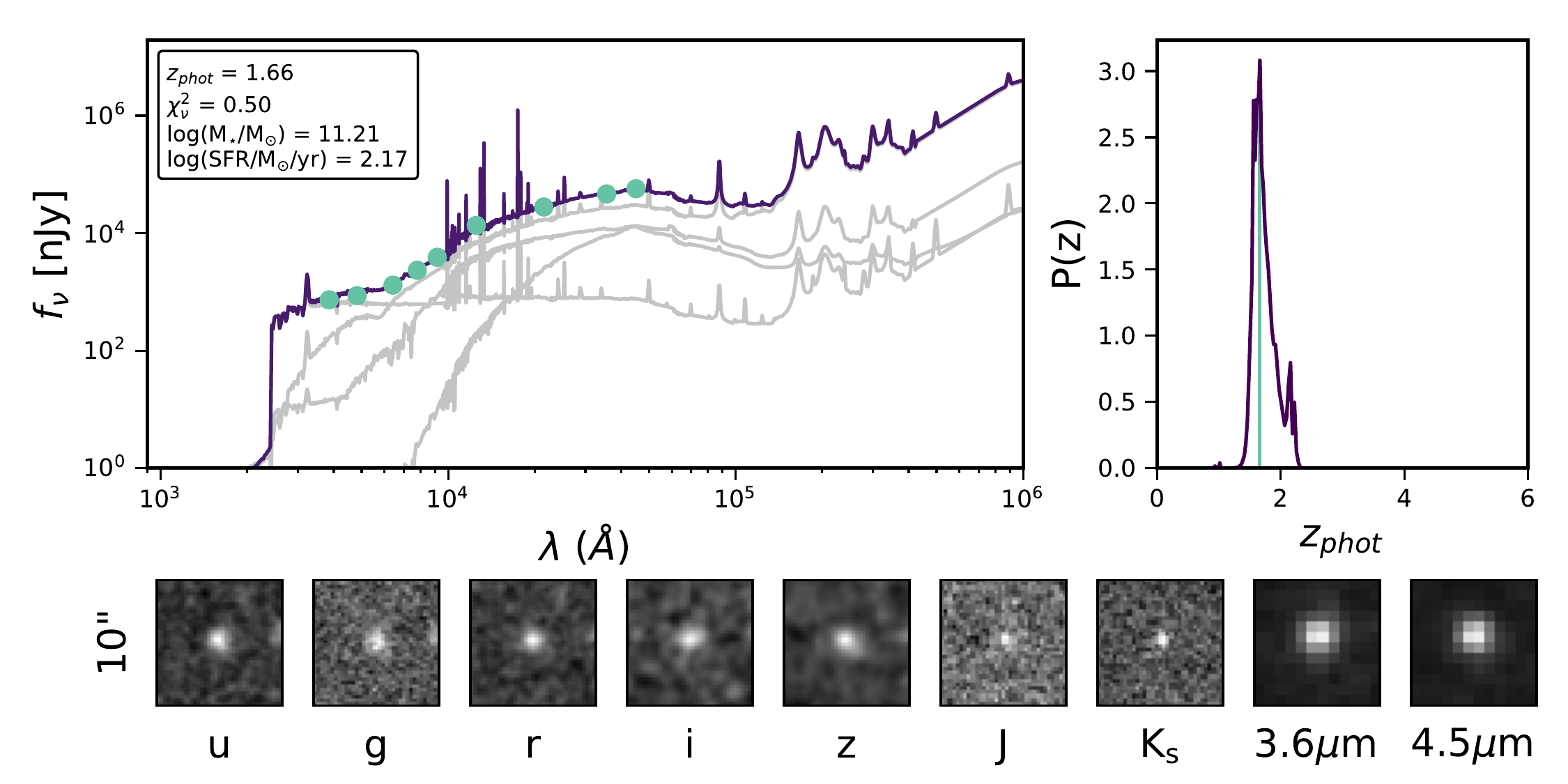} 
\includegraphics[width=\textwidth]{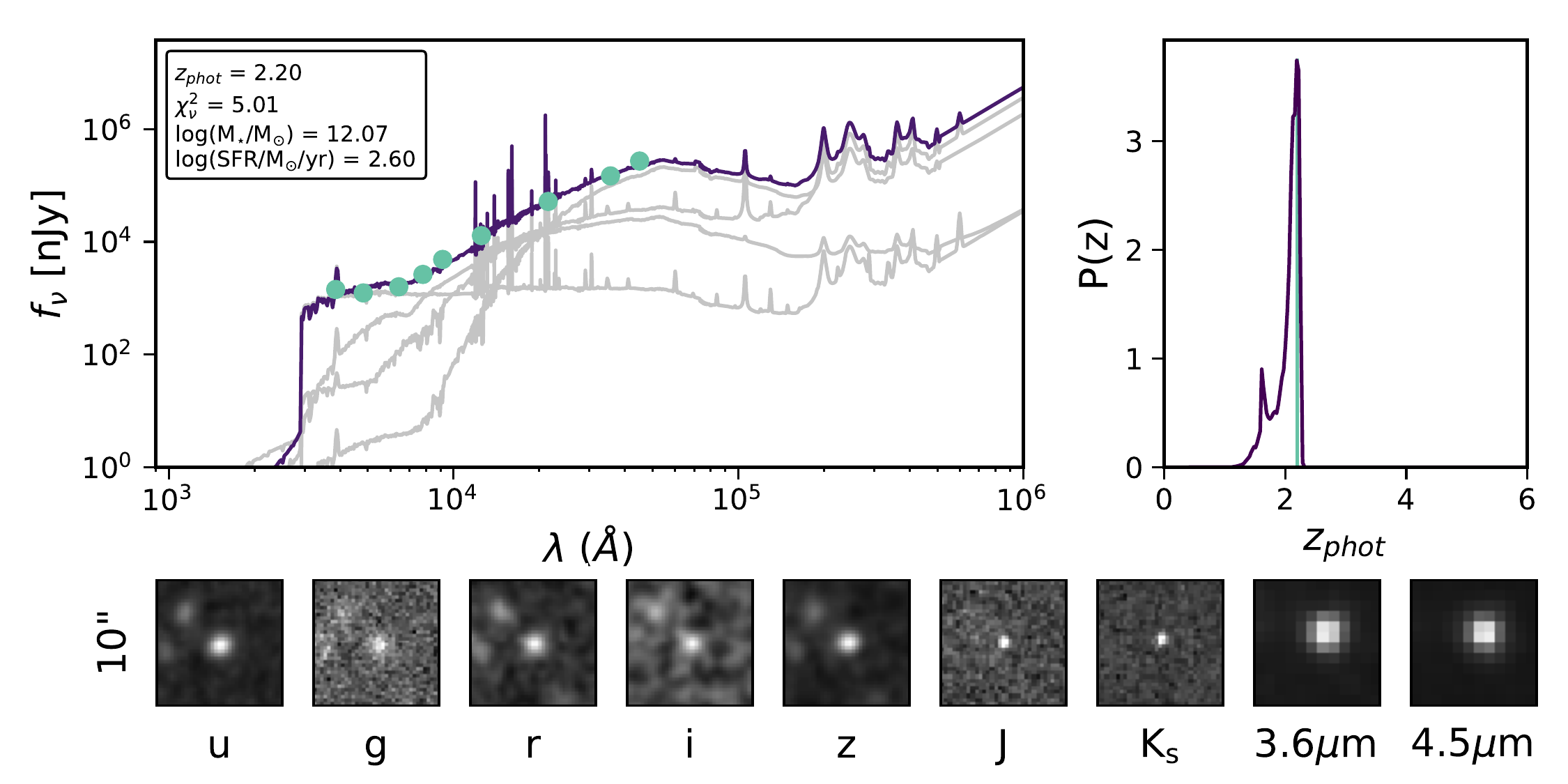} 
\caption{Example SED fits, photometric redshift probability distributions, and photometry for two $M_{\star} > 10^{11} M_{\odot}$ $1.5 < z < 2.5$ galaxies in our sample. Indicated in the inset box in the upper left panel are the object's best-fitting photometric redshift ($z_{\rm phot}$, the redshift at which $\chi^2$ is minimized), reduced $\chi^2$ ($\chi_{\nu}^2$) , stellar mass, and star-formation rate. The upper left panel shows the best-fitting SED template (purple) to the object's photometry (green points) with the set of templates (gray SED curves) added in non-negative linear combination to achieve the best-fitting SED. The upper right panel shows the photometric redshift probability distribution (purple curve), with the best-fitting redshift indicated by a vertical green line. Image cutouts show the photometry in all nine of our photometric bands. Each cutout is 10$\arcsec$ $\times$ 10$\arcsec$.
}
\label{sed_examples1}
\end{center}
\end{figure*}

\begin{figure*}
\begin{center}
\includegraphics[width=\textwidth]{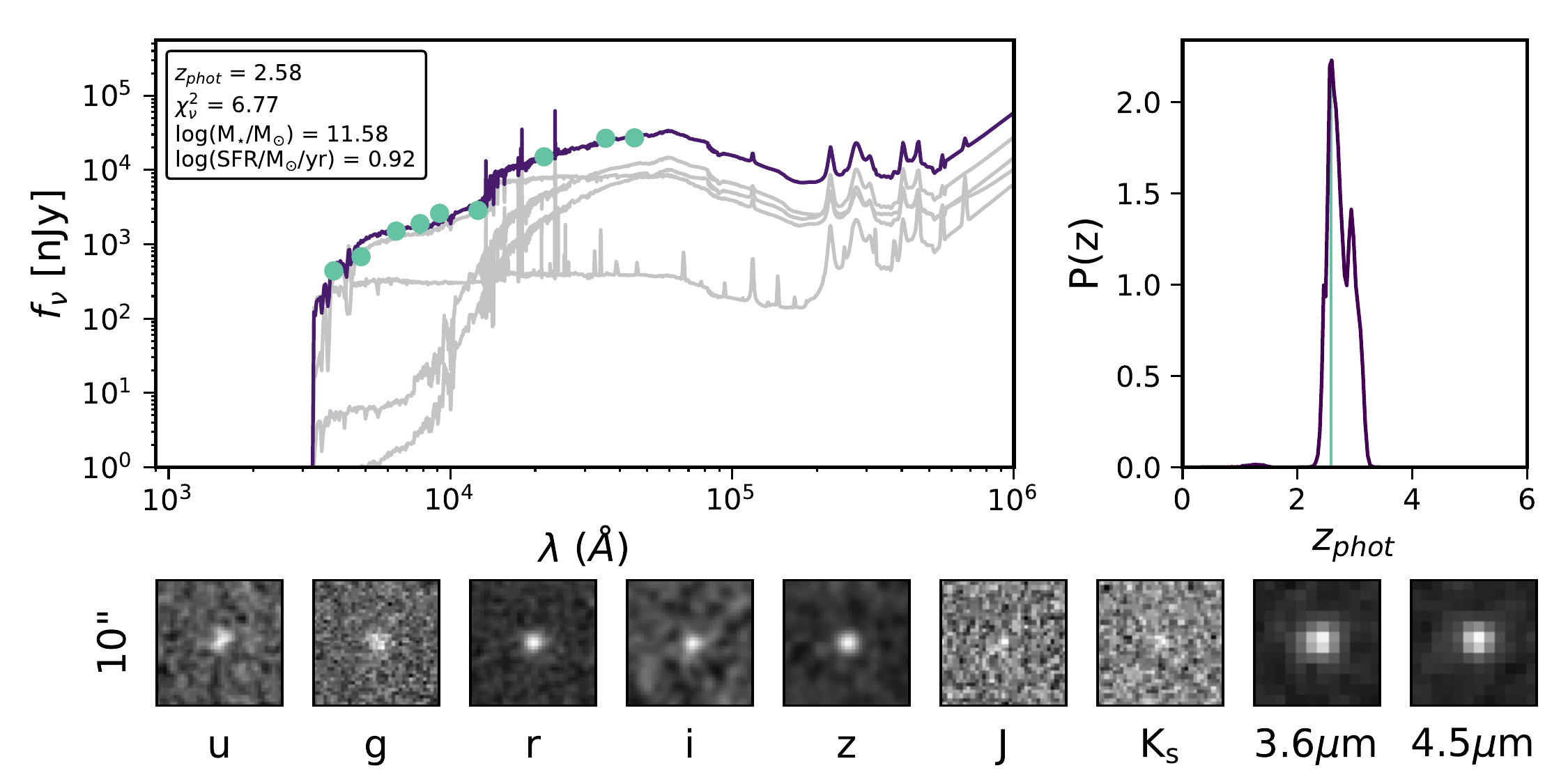} 
\includegraphics[width=\textwidth]{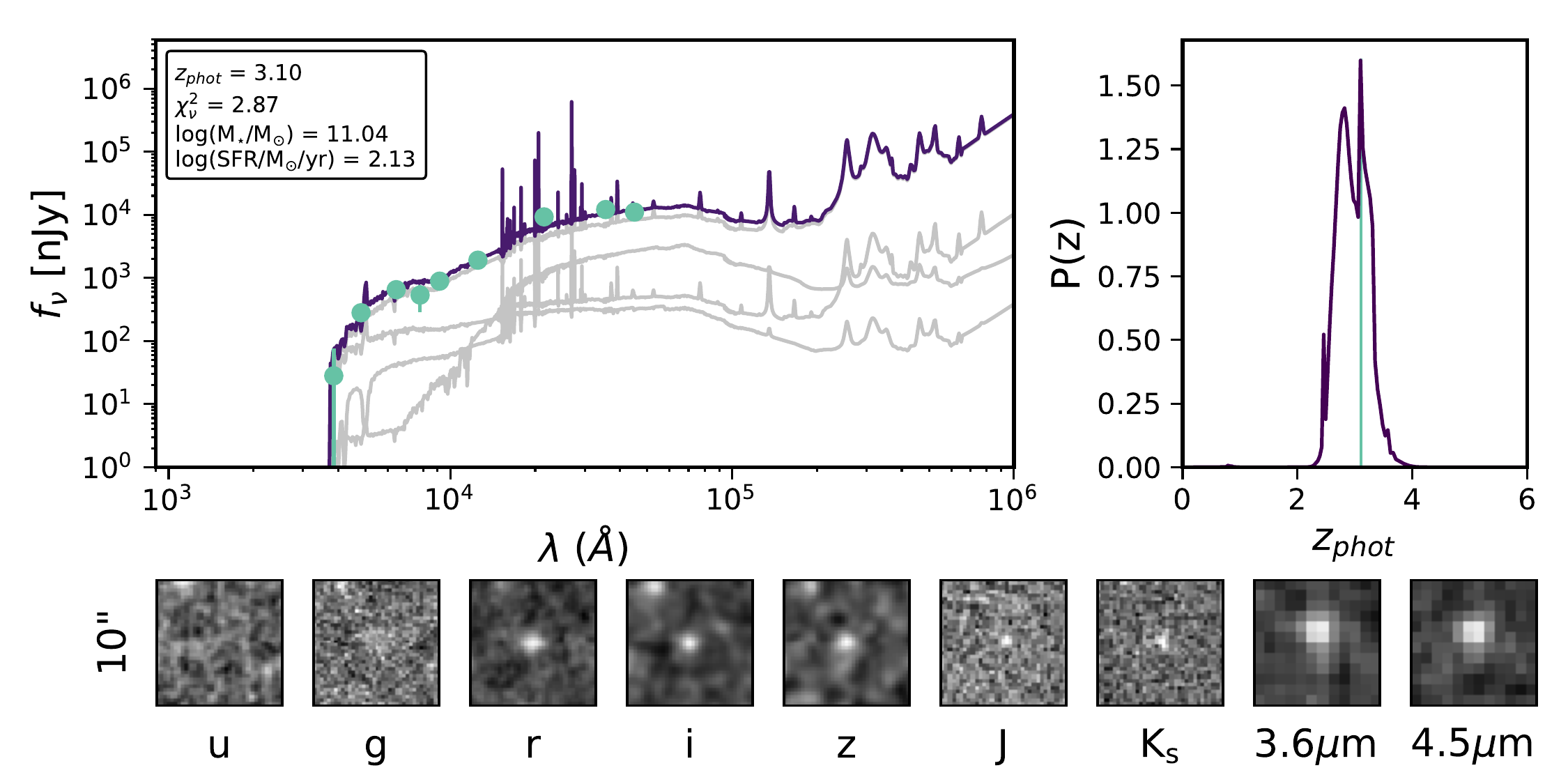} 
\caption{Same as Figure \ref{sed_examples1} for two $M_{\star} > 10^{11} M_{\odot}$  $2.5 < z < 3.5$ galaxies. }
\label{sed_examples2}
\end{center}
\end{figure*}

Photometric redshifts were obtained using EAZY-py\footnote{\url{https://github.com/gbrammer/eazy-py}}, an updated Python-based version of the template-based code EAZY \citep{Brammer2008}. EAZY-py fits a set of twelve Flexible Stellar Population Synthesis (FSPS) templates (\citealt{Conroy2009}, \citealt{Conroy2010}) in non-negative linear combination. The twelve default FSPS models included with EAZY-py span a wide range of galaxy types (star-forming, quiescent, dusty), with different realistic star-formation histories (SFH; bursty and slowly rising). The best-fit combination of templates is determined through $\chi^2$ minimization. This method is preferable when fitting a diverse sample of low- and high-redshift galaxies as it mitigates the bias to low-redshift objects that may occur when using single templates based primarily on low-redshift galaxies. The redshift distribution of galaxies in the SHELA footprint is shown in Figure \ref{zphot_hist}, and example spectral energy distribution (SED) fits and photometry of galaxies fit to have $M_{\star} > 10^{11} M_{\odot}$ and $1.5 < z < 3.5$ are shown in Figure \ref{sed_examples1} and Figure \ref{sed_examples2}. 

Properties such as photometric redshift, stellar mass, and star formation rate from SED fitting procedures inherently carry some uncertainties. We test the reliability of the SED fits by comparing photometric and spectroscopic redshifts later in this section, as well as investigating how well we can recover the properties of mock galaxies with known redshift, stellar mass, and star formation rate (see Section \ref{sec:fake_galaxies}).

Even with the careful construction of the EAZY-py template set, which is meant to represent a wide range of galaxy SED templates spanning quiescent and star-forming populations with a simple set of twelve templates, there is still the potential that these templates (or any combination thereof) are not perfect representations of the real galaxies in the SHELA footprint. These discrepancies appear in output SEDs as systematically higher or lower flux densities extracted from the best-fitting SED in a particular band compared to observed source photometry in that band. To combat this potential discrepancy and account for systematic offsets in the template set, we utilize a sample of 4,951 galaxies in the SHELA footprint that have spectroscopic redshifts from SDSS. These spectroscopically confirmed galaxies are used to compute the SED template-based magnitude offsets in each photometric band using the built-in EAZY-py magnitude offset function. For the sample of SDSS galaxies in the SHELA footprint (all z < 1), we find extremely small offsets in each band (Table \ref{mag_offset_table}), indicating that the EAZY-py method of fitting all templates in non-negative linear combination can produce SEDs that fit the observed population well without extensive adjustments. 

\begin{table}
\begin{center}
\caption{SED template-based magnitude offsets determined for each photometric band using low redshift galaxies in the SHELA footprint that have spectroscopic redshifts from SDSS. These offsets are self-consistently computed within EAZY-py and are applied to source photometry such that f$_{\rm{offset}}$ =  Offset $\times$ f$_{\rm{catalog}}$, where f$_{\rm{catalog}}$ is the flux density in the published catalog \citep{Wold2019}, ``Offset" is the SED template-based magnitude offset determined by EAZY-py, and f$_{\rm{offset}}$ is the object's flux after applying the magnitude offset. }
\label{mag_offset_table}
\begin{tabular}{ll}
\hline \hline 
Photometric Band		&Offset \\
\hline
DECam \textit{u}		&0.9923\\
DECam \textit{g}		&0.9908\\
DECam \textit{r}		&1.0000\\
DECam \textit{i}		&0.9820\\
DECam \textit{z}		&0.9823\\
VICS82 \textit{J}		&0.9968\\
VICS82 \textit{$K_s$}	&0.9916\\
IRAC 3.6 $\mu m$		&0.9987\\
IRAC 4.5 $\mu m$		&0.9845\\
\hline  
\end{tabular}
\end{center}
\end{table}

To quantify the photometric redshift error that results from the fits of SDSS-matched galaxies with EAZY-py, the normalized median absolute deviation \citep{Brammer2008} is computed:
\begin{eqnarray}
\sigma_{\rm NMAD} &=& 1.48 \times \rm{median}\left( \left| \frac{\Delta z - \rm{median}(\Delta z)}{1 + z_{spec}}  \right| \right).
\end{eqnarray}
With the previously computed magnitude offsets applied when fitting for photometric redshift, we find that $\sigma_{\rm NMAD}$ = 0.0377 for the $z < 1$ galaxies in our survey footprint that have spectroscopic redshifts from SDSS. This value of $\sigma_{\rm NMAD}$ is based on bright SDSS sources at low redshifts and may not be representative of the fainter and/or higher redshift galaxies in our sample\footnote{Using a preliminary internal data release from the HETDEX spectroscopic survey, we perform this same test of photometric redshift accuracy on a small sample of 16 galaxies from our $1.5 < z_{\rm phot} < 3.5$ sample that have HETDEX spectroscopic redshifts. For this small sample we find $\sigma_{\rm NMAD}$ = 0.168, indicating fair agreement between our photometric and spectroscopic redshifts.}. We therefore complement this SDSS-based test of photometric redshift recovery with a different test based on the recovery of photometric redshifts for mock galaxies modeled at higher redshifts (see Section \ref{sec:fake_galaxies}).

\subsubsection{Stellar Mass}
\label{sec:stellar_mass}
Stellar masses and other galaxy parameters, such as star-formation rate, for galaxies in the SHELA footprint are generated using EAZY-py. The FSPS templates (\citealt{Conroy2009}, \citealt{Conroy2010}) implemented in EAZY-py have gained popularity in recent years due to the options for implementing star-formation histories that are more realistic (e.g., bursty SFH, slowly rising and falling SFR) than the commonly used exponentially declining model (e.g., \citealt{Reddy2012}, \citealt{Leja2017}). We use the default EAZY-py FSPS models that are built with a \cite{Chabrier2003} initial mass function, \cite{KriekConroy2013} dust law, and solar metallicity. 

In order to compute the errors on stellar mass and other galaxy parameters, we draw 100 SEDs for each source from the best-fitting SED's template error function. This procedure ultimately gives 100 values of each galaxy parameter for every object. For each distribution of galaxy parameters, we assign the 50$^{\rm th}$ percentile of the galaxy parameter distribution to be the median-fit value, and the 16$^{\rm th}$ and 84$^{\rm th}$ percentiles as the lower and upper error bars, respectively. The median-fit is adopted throughout this work. 

To estimate stellar mass completeness, we follow the procedure of \cite{Pozzetti2010} and \cite{Davidzon2013} in which mass completeness is determined using the galaxies found in our survey. The backbone of this method is the premise that the mass completeness limit of a survey is the mass of the least massive galaxy that can be detected in a given bandpass with a magnitude equal to the limiting magnitude of the survey in that bandpass. At each redshift, we select the 20\% faintest objects and scale their mass (log($M_{\star, \rm EAZY-py}$)) such that their AB magnitude ($m$) equals the magnitude limit ($m_{\rm lim}$) of the survey in a given bandpass:
\begin{eqnarray}
\log(M_{\star, m=m_{\rm lim}}) = \log(M_{\star, \rm{EAZY-py}}) + 0.4(m - m_{\rm lim}).
\end{eqnarray}
After scaling the mass of the 20\% faintest objects, we find the 80$^{\rm th}$ percentile of the mass distribution and assign this value to be the 80\% mass completeness limit in each redshift bin. While mass completeness is usually estimated using a filter closest to the rest-frame $K_s$-band (in our case, the IRAC 4.5$\mu$m band), we choose to estimate the mass-completeness using the $r$-band. While the $r$-band does not directly trace the stellar mass buildup of a galaxy, without a strong enough detection in the $r$-band, it is unlikely that a source will have significant detections in enough filters to obtain a successful SED fit, without which we don't have redshift or stellar mass information. Using the $r$-band, with the 5$\sigma$ 80\% completeness of 24.8 AB mag reported by \cite{Wold2019}, we find that at the bounds of our redshift range of interest, $z$ = 1.5 (3.5), the 80\% mass completeness is log($M_{\star}$/$M_{\odot}$) = 10.89 (11.35). Regions of parameter space falling below the 80\% mass completeness limits for individual redshift ranges are indicated by gray shaded regions throughout this paper. 

In selecting our sample, we require detections with S/N $\geq$ 5 in r-band and IRAC 3.6 and 4.5$\mu$m (see Section \ref{sec_highConfidence}). With this in mind, we also performed this procedure using both IRAC bands with 5$\sigma$ 80\% completeness of 22.0 AB mag reported by \cite{Papovich2016} and found that in every redshift bin, the IRAC-based 80\% mass completeness limit is less than that found using the r-band. At the bounds of our redshift range of interest $z$ = 1.5 (3.5), the 80\% mass completeness is log($M_{\star}$/$M_{\odot}$) = 10.41 (10.89) for the 3.6$\mu$m band and log($M_{\star}$/$M_{\odot}$) = 10.41 (10.81) for the 4.5$\mu$m band. In using the r-band, we therefore report the most conservative estimate of the 80\% mass completeness limit for our sample.

\subsubsection{Selecting A High-Confidence Sample} 
\label{sec_highConfidence}
The ultimate goal of this work is to compile a sample of the most massive star-forming galaxies from $1.5 < z < 3.5$, however it is important to select a clean sample of galaxies at all masses in order to put the massive galaxy population into the overall context of galaxy evolution. In order to achieve this, we must select a sample of galaxies that are highly likely to be $1.5 < z < 3.5$ and fit with accurate stellar mass and related stellar population parameters. We require that each source in our ``high-confidence" sample meet several criteria in order to remain in the sample. 

At the most basic level, we require that the source is not identified as a star or AGN by SDSS. This requirement removes $\sim$2\% of objects from our catalog at all redshifts. Known luminous AGN are removed due to the lack of specific fitting procedures to account for AGN components in EAZY-py. It has been shown (e.g., \citealt{Salvato2011}, \citealt{Ananna2017}, Florez et al.\ in preparation) that without specifically accounting for emission from the AGN and the dusty torus, photometric redshifts and other derived quantities (stellar mass, star formation rate, dust extinction, etc.) are unreliable. With consideration to low luminosity AGN, we do not have a method for identifying these systems, however \cite{Salvato2011} showed that galaxies with low-luminosity AGN (defined in their study to have $F_{0.5-2keV}<8\times10^{-15}~cgs$) were adequately fit by normal galaxy templates without the specific implementation of AGN templates. Therefore, while low-luminosity AGN are not removed from our sample, we believe that they are reasonably fit with the SED fitting procedure used for our whole catalog. Additionally, we remove all galaxies with SDSS spectroscopic redshifts (all $z_{\rm spec} < 1$) from our sample. Only one SDSS $z_{\rm spec} < 1$ galaxy was wrongly fit to have photometric redshift $1.5 < z < 3.5$, and this one contaminant was removed.

Sources in this high confidence sample must have an SED fit that is constrained by six or more photometric bands and has $\chi_{\nu}^2<20$ for the best-fitting EAZY-py template combination. A particular band is used in the fit when $f_\nu>-90$ nJy (the EAZY-py ``not\_obs\_threshold", below which, a source is considered a non-detection) and there are no flags indicating a problem in that band. Specifically, we require that SExtractor external flags equal zero (no data quality issues), SExtractor internal flags are less than or equal to three (indicating a deblended source or source with no flags), and Tractor flags equal zero in each IRAC band (sources were successfully deblended). The requirement of six or more filters, which removes $\sim$15\% of objects from $1.5 < z < 3.5$, is chosen such that the SED fit is relatively well constrained. In cases where there are five or fewer filters, we found that even though a ``best-fit" template was fit by EAZY-py (sometimes with a low $\chi_{\nu}^2$ value), the same photometry could be reasonably fit by several different template combinations. 

Through visual inspection, we found that diffraction spikes and diffuse light from nearby stars or foreground galaxies can artificially inflate the IRAC flux for a given object, which is particularly troubling for our high-mass sample as these inflated IRAC fluxes result in unrealistically high stellar masses. In order to identify and remove these objects from our sample, we perform visual inspection of all objects fit to have $M_\star$ $>$ $10^{11}$$M_\odot$. Visual inspection is conducted three times for each source using the Zooniverse\footnote{\url{zooniverse.org}} interface. We ultimately find that the contamination fraction is $\sim$2\% for log($M_\star$/$M_\odot$) < 11.5 and rapidly increases to 100\% contamination at the highest masses (log($M_\star$/$M_\odot$) $\gtrsim$ 12.5) fit by EAZY-py, further emphasizing the importance of visual inspection in the high-mass galaxy regime. After flagging these contaminated sources, they are removed from our sample.

These initial quality cuts leave us with a sample of 158,879 galaxies at $1.5 < z < 3.5$ with robust photometric redshift measurements and obvious contaminants removed. In our redshift range of interest ($1.5 < z < 3.5$) it is imperative that objects have robust measurements in both IRAC filters in order to constrain the stellar mass. Over this entire redshift range, the only filters consistently redward of the Balmer break are VICS82 $K_s$ and IRAC 3.6 and 4.5$\mu$m (VICS82 $J$ is initially redward, but moves blue of the Balmer break in our $z = 2.5$ bin). As the VICS82 data do not cover the entirety of our field, we only place constraints on the IRAC data. To remove galaxies with unreliable stellar mass, we impose a S/N $\geq$ 5 requirement for an object to remain in our ``high confidence" sample. Following this cut (and previous quality cuts described above) we are left with a sample of 45,050 galaxies with robust measures of stellar mass at $1.5 < z < 3.5$, of which 17,609 (60.9\%) have $M_\star$ $>$ $10^{11}$$M_\odot$. 

Our science case focuses on star-forming galaxies, and we must remove quiescent galaxies in order to achieve a clean sample. To do this we impose two cuts, the first of which is a S/N $\geq$ 5 requirement in the r-band which reduces our sample size to 5,510 galaxies with $M_\star$ $>$ $10^{11}$$M_\odot$. Second, we require that galaxies in our sample have specific star formation rate sSFR > $10^{-11}~yr^{-1}$, following \cite{Fontana2009} and \cite{Martis2016} where sSFR =SFR/$M_\star$. This gives us a final sample of 5,352 massive ($M_\star$ $>$ $10^{11}$$M_\odot$) star-forming galaxies at 1.5 < z < 3.5, the largest uniformly selected sample of these objects to date. Applying the selection for star-forming galaxies removed $\sim$69\% of galaxies with $M_\star$ $>$ $10^{11}$$M_\odot$, which is consistent with results from \cite{Fontana2009} and \cite{Martis2016} that find 50-70\% of galaxies at the high mass end to be quiescent. 

Finally, we estimate the number of massive star-forming galaxies that we may be missing as a result of imposing the strict IRAC S/N $\geq$ 5 cut needed for robust estimates of stellar mass. If we take the stellar mass fit by EAZY-py at face value, we find that 511 galaxies with S/N < 5 in both IRAC filters have S/N $\geq$ 5 in the r-band and are fit to have $M_\star$ $>$ $10^{11}$$M_\odot$ at $1.5 < z < 3.5$. These 511 potentially real massive star-forming galaxies that were removed from our sample would constitute $\sim$9.5\% of our final sample of massive star-forming galaxies at these redshifts. 

\subsection{Testing EAZY-py With Mock Galaxies}
\label{sec:fake_galaxies}
Due to the uncertainties that come with SED fitting, it is important to confirm that the SED fitting procedure and models used (in our case, EAZY-py with FSPS models) can correctly recover galaxy redshifts and parameters for galaxies with our filter set. Testing redshift recovery is difficult at $z > 1$ due to the limited number of spectroscopic redshifts available. Galaxy spectroscopic redshifts in the SHELA field come solely from SDSS and are for $z < 1$ galaxies. Future optical spectroscopy from HETDEX will provide spectroscopic redshifts for Ly$\alpha$ emitters within 1.9 < $z$ < 3.5, and these will provide a sample to test the accuracy of photometric redshift recovery at $z > 1$. Since these samples do not currently exist, we rely on our sample of mock galaxies to test our photometric redshift recovery for $1.5 < z < 3.5$ galaxies.

We begin by constructing a sample of mock galaxies (Acquaviva, V., Private Communication) built from an expanded version of GALAXEV \citep{BruzualCharlot2003}. Several thousand $z \sim 0$, $M_{\star} \simeq 10^{10} M_{\odot}$ models were generated to span several initial mass functions (IMF; Chabrier, Salpeter, and Kroupa), dust laws (Calzetti and Milky Way), and star formation histories (exponentially declining, delayed exponential, constant, and linearly increasing) for a large grid of age, e-folding time ($\tau$), extinction, and metallicity values. This base set of models is then redshifted to $z = 0.001 - 6$ with $\Delta$$z = 0.2$ and scaled in mass to $M_{\star} \simeq [1\times10^{10}, 5\times10^{10}, 1\times10^{11}, 5\times10^{11}, 1\times10^{12}] M_{\odot}$. Ultimately, this passive model scaling leads to $\sim$ 2.4 million mock galaxies. Finally, from each full mock galaxy spectrum, we extract photometry in the filters available in the SHELA footprint (DECam \textit{ugriz}, VICS82 \textit{J}, $K_s$, and \textit{Spitzer}-IRAC 3.6 and 4.5$\mu m$).

Assigning errors to the extracted mock galaxy photometry is a three step process, completed separately for each photometric band of every mock galaxy. In the first step, we find an object in the SHELA footprint ``high-confidence" sample (see Section \ref{sec_highConfidence}) with similar magnitude (within $\Delta$mag = 0.2) in a particular photometric band to the mock galaxy's magnitude in that band. Using all of the SHELA footprint galaxies found to be in this magnitude bin, we construct a distribution of flux errors associated with those real SHELA footprint galaxies and an error value is drawn from that distribution. In the second step, the value of the flux for a given mock galaxy in a particular photometric band is scattered. This is done by constructing a Gaussian centered on the original mock galaxy's flux with $\sigma$ equal to the error value drawn in the first step of this process. The scattered flux in that photometric band is then drawn from the Gaussian. Scattering is done to re-create the random error associated with extracting photometry from images in the SHELA footprint. Finally, in the third step, we repeat the procedure of step one using the scattered mock galaxy magnitude found in step two as the magnitude around which we search for SHELA footprint galaxies and select a new error bar. This three step procedure produces a scattered flux with realistic errors for each photometric band of every mock galaxy. 

\begin{figure*}
\begin{center}
\includegraphics[width=\textwidth]{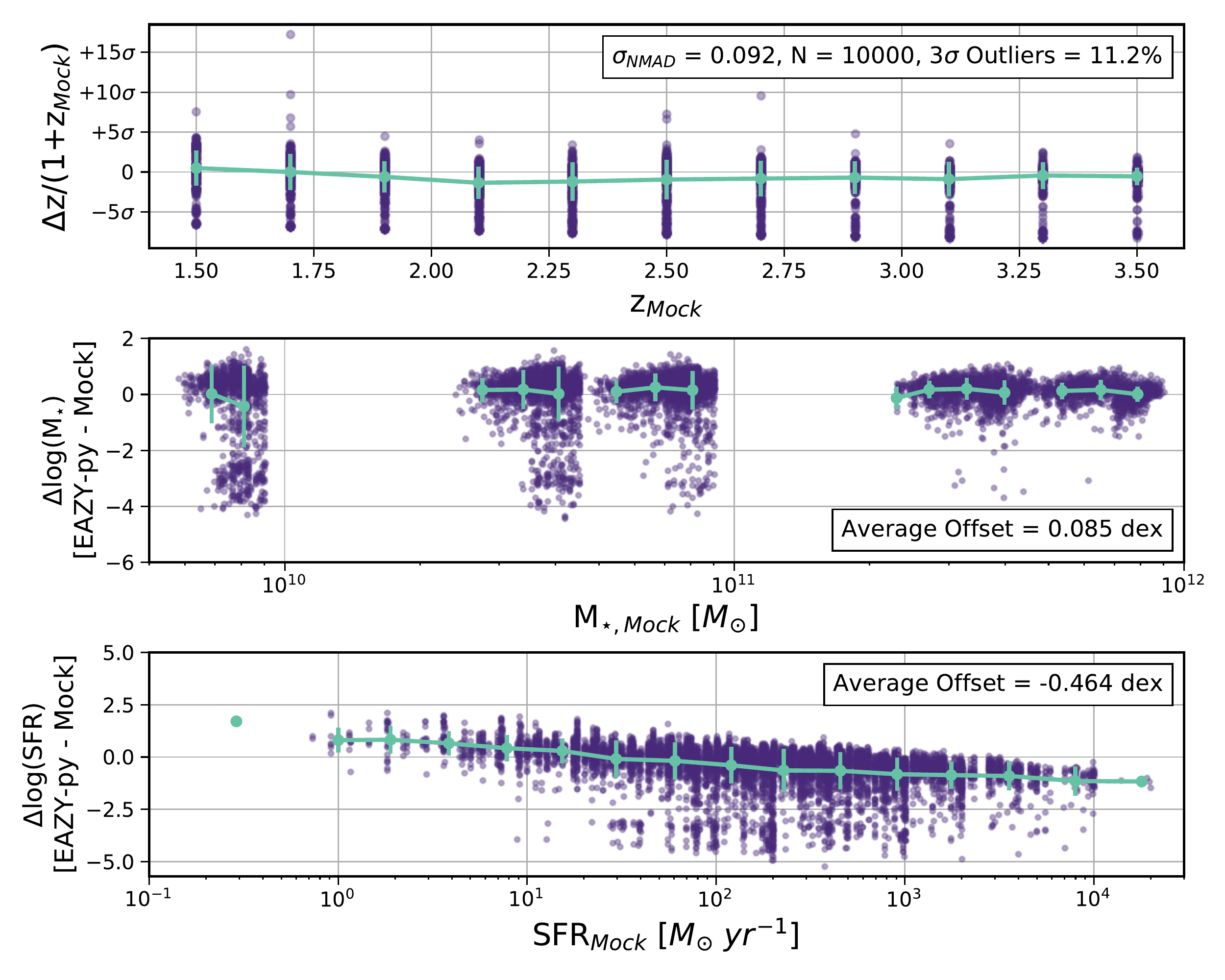}
\caption{Comparison of redshift (top panel), stellar mass (middle panel), and SFR (bottom panel) between EAZY-py output and the input values for 10,000 randomly selected mock galaxies. Error bars span the 68$\rm th$ percentile of the data in each bin, which corresponds to an average dispersion in redshift of 0.19 $\Delta z$/(1+$z_{\rm Mock}$), average dispersion in stellar mass of 0.58 dex, and average dispersion in SFR of 0.69 dex. We find that redshift recovery error ($\sigma_{\rm NMAD}$ = 0.092) is higher than that found using a low redshift sample of SHELA footprint galaxies with SDSS spectroscopic redshifts, however this increased error is expected. On average, we find that EAZY-py may over-predict stellar mass by $\sim$0.09 dex and under-predict SFR by $\sim$0.46 dex when comparing the output values from EAZY-py to those associated with the mock galaxies. There are no data points lying behind the caption boxes.}
\label{vivi_galaxies}
\end{center}
\end{figure*}

It is important to note that this procedure of passively scaling mock galaxy spectra constructed in various ways and adding error bars is \textit{not} akin to constructing a mock catalog, and therefore, does not represent a realistic distribution of galaxies. Rather, for this test we simply require a diverse set of mock galaxies with known properties (e.g., redshift, stellar mass, SFR) to test the ability of our SED fitting procedure to recover these properties. Our main interest for this work lies with galaxies at $1.5 < z < 3.5$ and our comparison here will focus on mock galaxies falling in this redshift range. In order to test EAZY-py, we run a randomly selected sample of 10,000 mock galaxies through EAZY-py and compare their output redshift, stellar mass, and SFR to the ``truth" associated with that mock galaxy. This random subset of mock galaxies contains models spanning the full range of IMFs, dust laws, SFHs, ages, e-folding times, extinction, and metallicity values used to construct our mock galaxy sample. 

The primary goal of this exercise of fitting mock galaxies is to determine if the SED fitting procedure used on real SHELA footprint galaxies can adequately recover the redshift, stellar mass, and star formation rate for the mock sources. Extreme discrepancies here in this somewhat idealized sample would be cause for concern with the use of EAZY-py to fit the sources in the SHELA footprint. 

We find that EAZY-py does a good job of recovering the redshift of the sources in the redshift range $1.5 < z < 3.5$, with a redshift error of $\sigma_{\rm NMAD}$ = 0.092 and 11.2\% 3$\sigma$ outliers (top panel of Figure \ref{vivi_galaxies}). Extreme outliers are generally mock galaxies with featureless SEDs or those fit to have highly multi-peaked photometric redshift distributions where there are several probable photometric redshift fits, but the redshift found by EAZY-py to have the lowest $\chi^2$ is not the true redshift for that object. Additionally, we find that EAZY-py, on average, slightly overestimates stellar mass by $\sim$0.09 dex and underestimates SFR by $\sim$0.46 dex (middle and bottom panels of Figure \ref{vivi_galaxies}, respectively). Systematic offsets may be partially attributed to differences between the BC03 models used to construct the mock galaxies and the FSPS models used in the EAZY-py SED fitting. These systematic offsets may also have contributions from the non-negative linear combination fitting method used in EAZY-py. 

We emphasize that these tests on mock galaxies are simply meant to demonstrate that EAZY-py can adequately recover the redshift, stellar mass, and star formation rate for a set of mock galaxies with widely different underlying SEDs. We do not to make any corrections to our data or theoretical models using the results of our mock galaxy tests because the sample of mock galaxies does not have the same relative distribution of different types of galaxies as our observational catalog. Instead, when needed, we use the average stellar mass error in each redshift bin based on SED fits performed on our galaxy sample (see Section \ref{sec:stellar_mass} for details) as a measure of the uncertainty on the stellar mass.

\section{Results}
\label{sec:smf}
\subsection{Massive Star-Forming Galaxies in the SHELA Footprint}
\begin{figure*}
\begin{center}
\includegraphics[width=\textwidth]{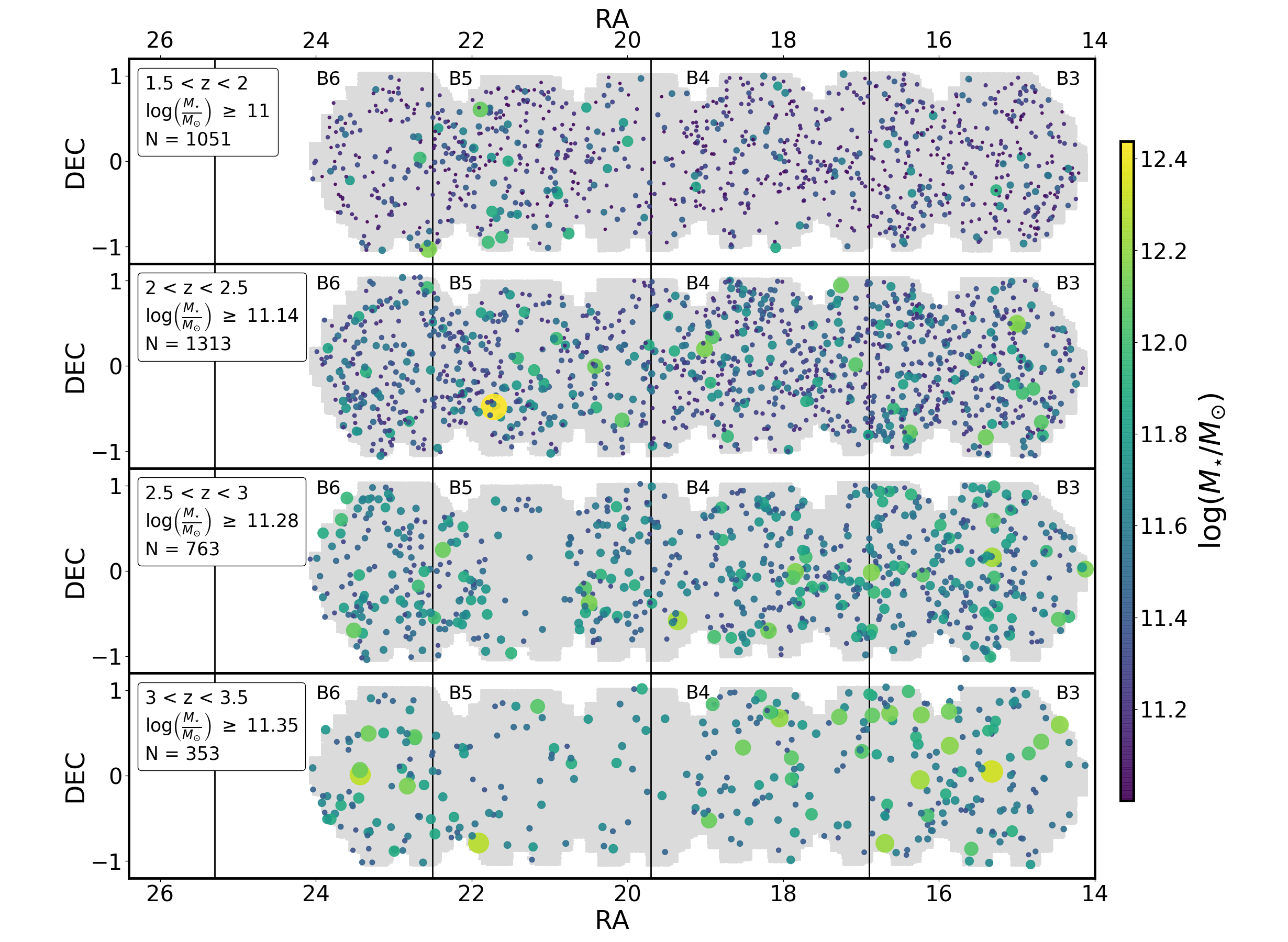} 
\caption{Map of the SHELA footprint (gray) with massive star-forming galaxies shown in color. The color and size of each point corresponds to the mass of the galaxy, with brighter, larger points representing more massive galaxies. Panels from top to bottom increase in redshift by $\Delta$$z = 0.5$ from $z = 1.5$ to $z = 3.5$. The lowest redshift bin ($1.5 < z < 2$) contains galaxies with $M_\star$ $>$ $10^{11}$$M_\odot$, while the higher redshift bins ($2 < z < 3.5$) contain galaxies more massive than the 80\% mass completeness limit determined for each redshift bin in Section \ref{sec:stellar_mass}. The minimum mass in each redshift bin and the number of galaxies above that mass threshold are shown in the upper left corner of each panel. Vertical lines represent the boundaries of the SHELA footprint tiles (B3, B4, B5, and B6) discussed in Section \ref{sec:data}, and tile names are indicated in each panel. The number per unit area of massive galaxies ($M_\star$ $>$ $10^{11}$$M_\odot$) at $1.5 < z < 3.5$ is quite low ($\sim$310/deg$^2$), emphasizing the benefit of our large area SHELA footprint in providing a statistically significant sample (5,352 with $M_\star$ $>$ $10^{11}$$M_\odot$) of massive star-forming galaxies.}
\label{massive_gal_map}
\end{center}
\end{figure*}

\begin{figure*}
\begin{center}
\includegraphics[width=\textwidth]{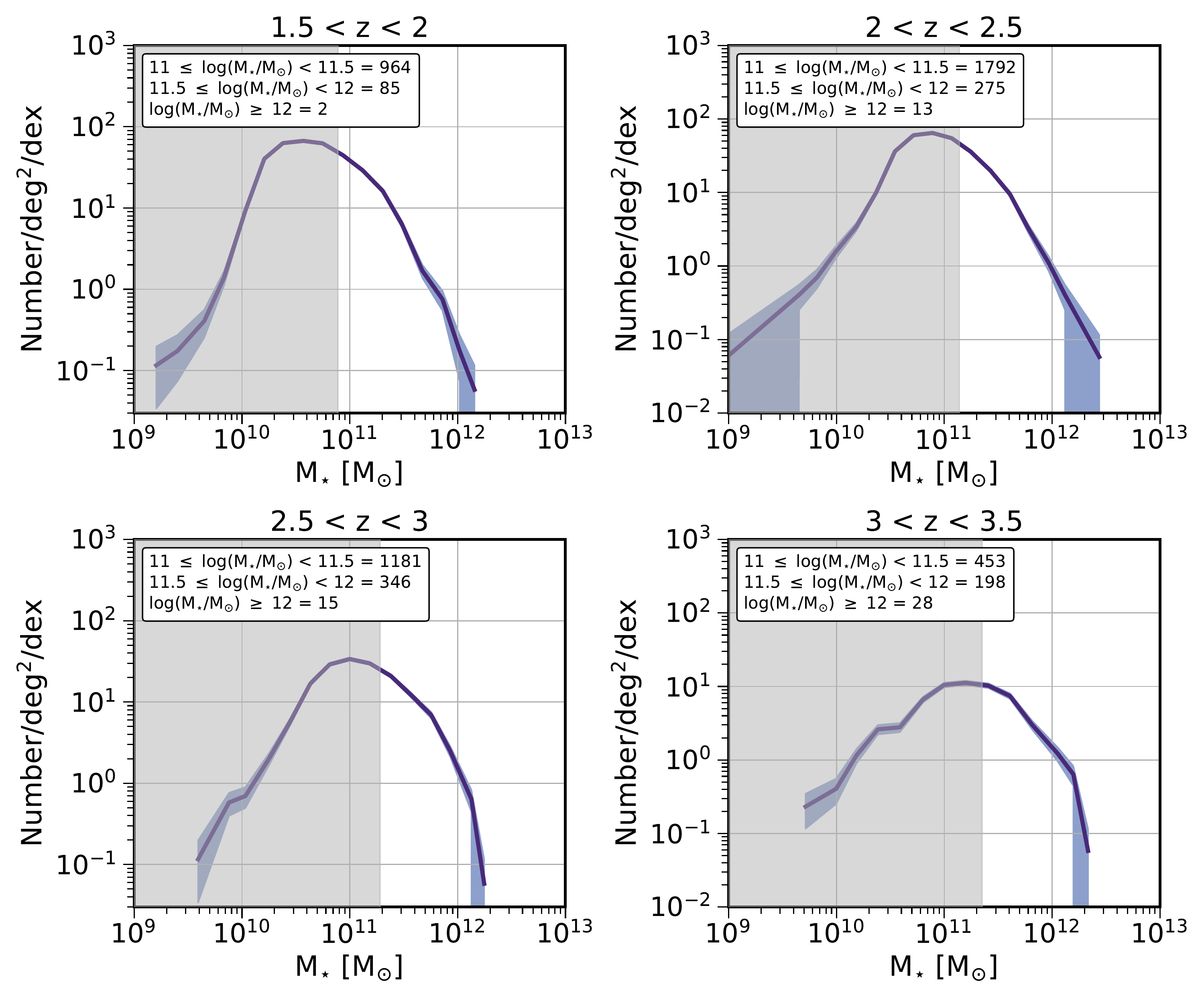}
\caption{The number of star-forming galaxies per square degree for $1.5 < z < 3.5$ galaxies in the SHELA footprint. The dark solid lines correspond to the number of sources found at each mass, while the light purple regions represent Poisson errors. The gray shaded regions represent masses below the 80\% mass completeness limit for each redshift bin. The total number of galaxies that we find in the high stellar mass regime are indicated in the inset box for each redshift bin.}
\label{npersqdeg}
\end{center}
\end{figure*}

We find that the sample of massive star-forming galaxies in the SHELA footprint is distributed across the entirety of the $\sim$17.2 deg$^2$ area (Figure \ref{massive_gal_map}). While this is not unexpected, it further highlights the necessity of a large area survey with a uniformly selected sample when compiling statistically significant populations of sparse objects. With our large survey footprint, we can provide measures of the number of objects per square degree in specific mass and redshift bins, which may prove useful in the design of future surveys. The total number of star-forming galaxies, number per square degree, and number density found in several mass and redshift bins is detailed in Table \ref{tab:num_obj}, while Figure \ref{npersqdeg}  shows the number per square degree as a function of mass in small redshift bins with $\Delta$$z = 0.5$ from $z = 1.5$ to $z = 3.5$. \textit{In all, we find 962 star-forming galaxies with log($M_\star$/$M_\odot$) $>$ 11.5, the largest uniformly selected sample of these objects found to date from $1.5 < z < 3.5$}. 

Intuitively, one would expect that the number of extremely massive galaxies (for the full star-forming and quiescent population) would increase towards present day, such that there should be more massive galaxies in our $1.5 < z < 2$ bin than there are in the $3 < z < 3.5$ bin. From Figure \ref{npersqdeg} and Table \ref{tab:num_obj}, we find that the number of the most massive ($M_\star$ $>$ $10^{12}$$M_\odot$) star-forming galaxies decreases toward our lowest redshift bin, as is expected for star-forming massive galaxies at this epoch. For star-forming galaxies with $10^{11}$$M_\odot$ $<$ $M_\star$ $<$ $10^{12}$$M_\odot$ we find that the number of massive star-forming galaxies increases from $z = 3.5$ to $z \sim 2$, and begins to decrease in the $1.5 < z < 2$ bin. Previous studies (e.g., \citealt{Fontana2009}, \citealt{Martis2016}) have shown that 50 - 70\% of massive ($M_\star$ $>$ $10^{11}$$M_\odot$) galaxies  are quiescent by $z \sim 1.5$. 

The SHELA footprint star-forming galaxy stellar mass function is computed using the $1/V_{\rm max}$ method \citep{Schmidt1968} following the procedure of \cite{Weigel2016}. We begin by finding the 95\% mass completeness limit as a function of redshift by repeating our mass completeness procedure described in Section \ref{sec:stellar_mass} with the 95\% r-band completeness limit of r = 23.5 AB mag from \cite{Wold2019}. This allows us to determine the redshift at which our data are 95\% complete for a given stellar mass. We bin our sample by photometric redshift into four redshift bins from $1.5 < z < 3.5$ with $\Delta z_{\rm bin} = 0.5$ and further bin by stellar mass within these redshift bins. The $1/V_{\rm max}$ term is then determined for each mass bin based on the maximum redshift at which that mass bin is 95\% complete ($z_{\rm max}$). In the case that $z_{\rm max}$ is greater than the maximum redshift of a particular redshift bin, we assign $V_{\rm max}$ to equal the comoving volume of that redshift bin. If $z_{\rm max}$ lies within the redshift bin of interest, then the $V_{\rm max}$ term is simply the comoving volume between the minimum redshift of that bin and $z_{\rm max}$. Finally, if $z_{\rm max}$ is less than the minimum redshift of a particular redshift bin we assign $V_{\rm max}$ to be the minimum of the comoving volume from z = 0 to $z_{\rm max}$ and the comoving volume for our redshift bin of interest. 

Our empirical star-forming galaxy stellar mass function, shown in Figures \ref{smf_edd}, \ref{smf_v_lit}, and \ref{smf_v_theory} as dark purple points, spans masses from log($M_\star$/$M_\odot$) $\sim$ 9 to log($M_\star$/$M_\odot$) $\sim$ 12.5. We find that at the high mass end, our stellar mass function ($\Phi$) is steeply declining, spanning values of $\Phi$ as large as $\sim$$10^{-4}$ Mpc$^3$/dex and as small as $\sim$5$\times$$10^{-8}$ Mpc$^3$/dex. Errors on our empirical stellar mass function are entirely from Poisson errors as the large area of our survey renders errors from cosmic variance negligible. In Section \ref{sec:eddington_bias}, we estimate the effect that Eddington bias \citep{Eddington1913} has on our measured stellar mass function. In Section \ref{sec:smf_v_lit} we compare our empirical star-forming galaxy stellar mass function to previous observational results, and in Section \ref{sec:smf_v_theory} we compare with results from theoretical models. 

\begin{table*}
\begin{center}
\caption{(1) Redshift bin for the full redshift range of our sample and divided into smaller bins of $\Delta$$z = 0.5$. (2) Total number of star-forming massive galaxies, number per square degree, and number density of galaxies found in our sample split into smaller mass bins (3-5). When comparing the number of galaxies found to have log($M_{\star}$/$M_{\odot}$) $>$ 11.5 (columns 4 and 5) to the works of \protect\cite{Muzzin2013}, \protect\cite{Ilbert2013}, and \protect\cite{Tomczak2014}, we find an order of magnitude more star-forming galaxies than these previous small-area studies. Errors represent Poisson uncertainties.}
\label{tab:num_obj}
\begin{tabular}{ccccc}
\hline \hline 
Redshift		&Sample Statistic &11 $\leq$ log($\frac{M_{\star}}{M_{\odot}}$) $<$ 11.5	& 11.5 $\leq$ log($\frac{M_{\star}}{M_{\odot}}$) $<$ 12	&log($\frac{M_{\star}}{M_{\odot}}$) $\geq$ 12 \\
(1)	&(2)	&(3)	&(4)	&(5)\\
\hline
				& Total N		&4390 $\pm$ 66.25									&904	 $\pm$ 30.07									&58 $\pm$ 7.62\\
$1.5 < z < 3.5$		&N/deg$^2$	&254.73 $\pm$	3.84									&52.46  $\pm$	1.74									&3.37 $\pm$ 0.44\\	
				&N/Mpc$^3$	&1.01 $\times$ $10^{-5}$	$\pm$ 1.52 $\times$ $10^{-7}$		&2.08 $\times$ $10^{-6}$ $\pm$ 6.91 $\times$ $10^{-8}$		&1.33 $\times$ $10^{-7}$ $\pm$ 1.75 $\times$ $10^{-8}$\\
\hline \hline
				& Total N		&964 $\pm$ 31.05									&85 $\pm$ 9.22									&2 $\pm$ 1.41\\
$1.5 < z < 2$		&N/deg$^2$	&55.94 $\pm$ 1.8									&4.93 $\pm$ 0.53									&0.12 $\pm$ 0.08\\	
				&N/Mpc$^3$	&9.19 $\times$ $10^{-6}$ $\pm$ 2.96 $\times$ $10^{-7}$		&8.10 $\times$ $10^{-7}$ $\pm$ 8.79 $\times$ $10^{-8}$		&1.91 $\times$ $10^{-8}$ $\pm$ 1.35 $\times$ $10^{-8}$\\
\hline
				& Total N		&1792 $\pm$ 42.33 									&275	 $\pm$ 16.58									&13 $\pm$ 3.61\\
$2 < z < 2.5$		&N/deg$^2$	&103.98 $\pm$ 2.46									&15.96 $\pm$ 0.95									&0.75 $\pm$ 0.21\\	
				&N/Mpc$^3$	&1.61 $\times$ $10^{-5}$ $\pm$ 3.81 $\times$ $10^{-7}$		&2.48 $\times$ $10^{-6}$ $\pm$ 1.49 $\times$ $10^{-7}$		&1.17 $\times$ $10^{-7}$ $\pm$ 3.25 $\times$ $10^{-8}$\\
\hline
				& Total N		&1181 $\pm$ 34.37									&346 $\pm$ 18.60									&15 $\pm$ 3.87\\
$2.5 < z < 3$		&N/deg$^2$	&68.53 $\pm$ 1.99									&20.08 $\pm$ 1.08									&0.87 $\pm$ 0.22\\	
				&N/Mpc$^3$	&1.06 $\times$ $10^{-5}$ $\pm$ 3.09 $\times$ $10^{-7}$		&3.11 $\times$ $10^{-6}$ $\pm$ 1.67 $\times$ $10^{-7}$		&1.35 $\times$ $10^{-7}$ $\pm$ 3.49 $\times$ $10^{-8}$\\
\hline
				& Total N		&453	 $\pm$ 21.28									&198	 $\pm$ 14.07									&28 $\pm$ 5.29\\
$3 < z < 3.5$		&N/deg$^2$	&26.29 $\pm$ 1.24									&11.49 $\pm$ 0.82									&1.62 $\pm$ 0.31\\	
				&N/Mpc$^3$	&4.19 $\times$ $10^{-6}$ $\pm$ 1.97 $\times$ $10^{-7}$ 		&1.83 $\times$ $10^{-6}$ $\pm$ 1.30 $\times$ $10^{-7}$		&2.59 $\times$ $10^{-7}$ $\pm$ 4.89 $\times$ $10^{-8}$\\
\hline
\end{tabular}
\end{center}
\end{table*}

\subsection{Eddington Bias}
\label{sec:eddington_bias}
Eddington bias is the phenomenon in which, due to photometric error and errors in estimating photometric redshift, lower mass galaxies may scatter into a higher mass bin (\citealt{Eddington1913}). This is particularly important to consider at the high-mass end of the stellar mass function, where the steep slope and low number statistics compared to intermediate masses are easily influenced by a small number of low mass interlopers. In our mass range of interest ($M_\star$ $>$ $10^{11}$$M_\odot$), the primary error which causes a low-mass galaxy to be placed in a high-mass bin is fitting a low redshift galaxy at a high redshift. Errors on stellar mass estimates can also contribute in the steeply declining portion of the stellar mass function (Kawinwanichakij et al.\ in preparation), however stellar mass errors are incorporated into our SED fitting procedure as described in \ref{sec:stellar_mass} and we do not further implement them here for the purposes of estimating Eddington Bias. 

\begin{figure*}
\begin{center}
\includegraphics[width=\textwidth]{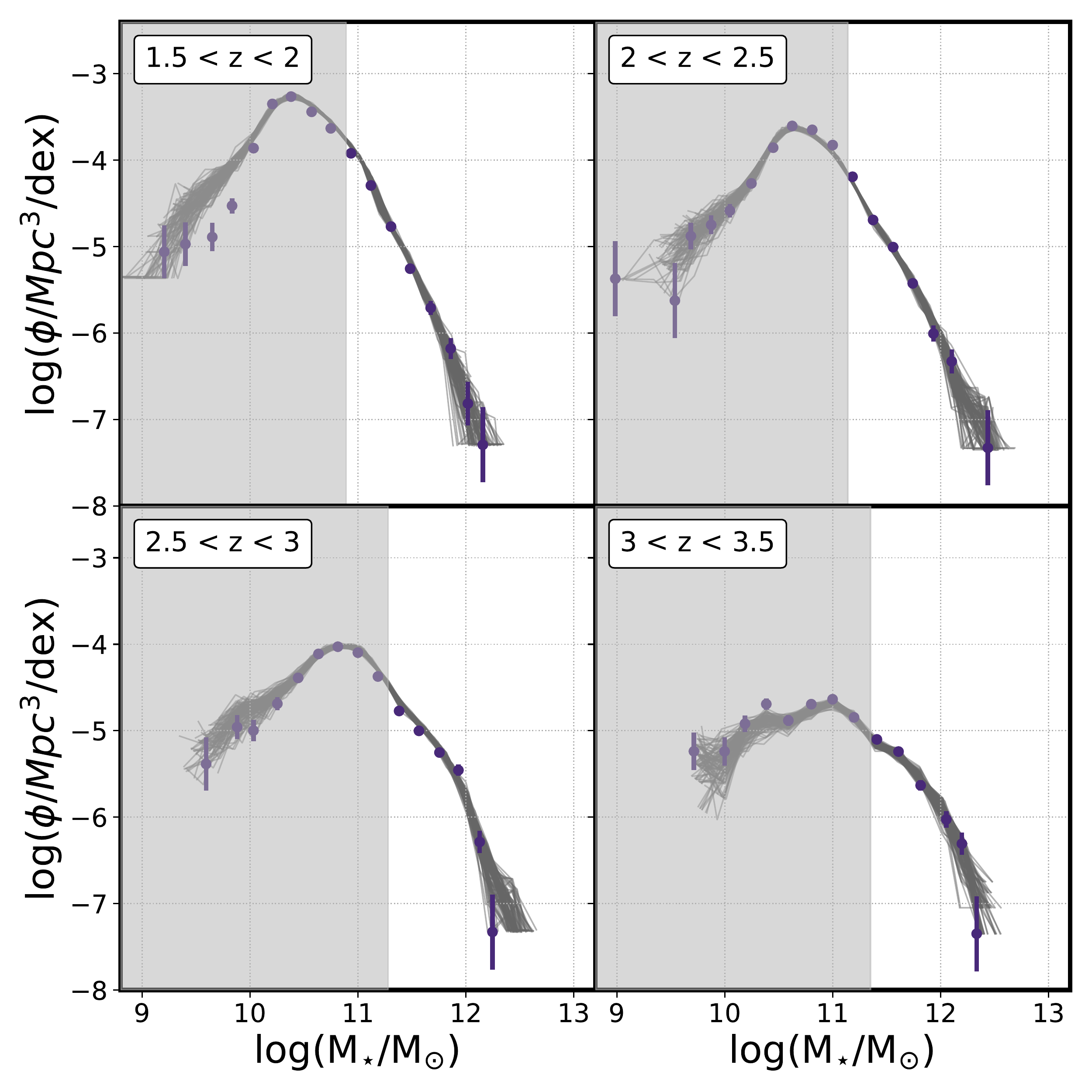} 
\caption{Purple points represent the star-forming galaxy stellar mass function for $1.5 < z < 3.5$ galaxies in the SHELA footprint, where error bars represent Poisson errors. Gray lines are stellar mass functions generated from 100 realizations of the catalog in which photometric redshifts are drawn from each object's photometric redshift probability distribution. These 100 realizations are used to estimate the effects of Eddington bias, as outlined in Section \ref{sec:eddington_bias}. The gray shaded regions represent masses below the 80\% mass completeness limit for each redshift bin. We find that scatter is introduced to the stellar mass function when the catalog is perturbed, however results remain consistent with those from the original catalog.}
\label{smf_edd}
\end{center}
\end{figure*}

To approximate the effects of Eddington bias on our results, we generate 100 realizations of our catalog in which we draw a new photometric redshift value from each object's redshift probability distribution. We perform SED fitting at the selected photometric redshift to generate 100 realizations of our catalog. We note that this procedure is performed for objects in the catalog at all redshifts such that objects originally outside $1.5 < z < 3.5$ may scatter into this range, while those originally inside this redshift range may scatter out. Finally, we construct stellar mass functions for each of the 100 realizations of the catalog (Figure \ref{smf_edd}). These stellar mass function realizations are constructed in the same manner as the ``high-confidence" star-forming galaxy sample (see Section \ref{sec_highConfidence}) used to construct our empirical stellar mass function (purple points on Figure \ref{smf_edd}). 

We find that these 100 realizations lead to a scatter in the high-mass end of the stellar mass function of 0.04 - 0.8 dex compared to the empirical SHELA footprint result. We discuss how Eddington bias affects comparisons of our empirical star-forming galaxy stellar mass function with previous stellar mass function results from the literature in Section \ref{sec:smf_v_lit} and from theoretical models in Section \ref{sec:smf_v_theory}, however we do not make any corrections to our empirical stellar mass function based on the Eddington bias estimates.

\subsection{Comparing SHELA Footprint SMF to Previous Observational Results}
\label{sec:smf_v_lit}
\begin{figure*}
\begin{center}
\includegraphics[width=\textwidth]{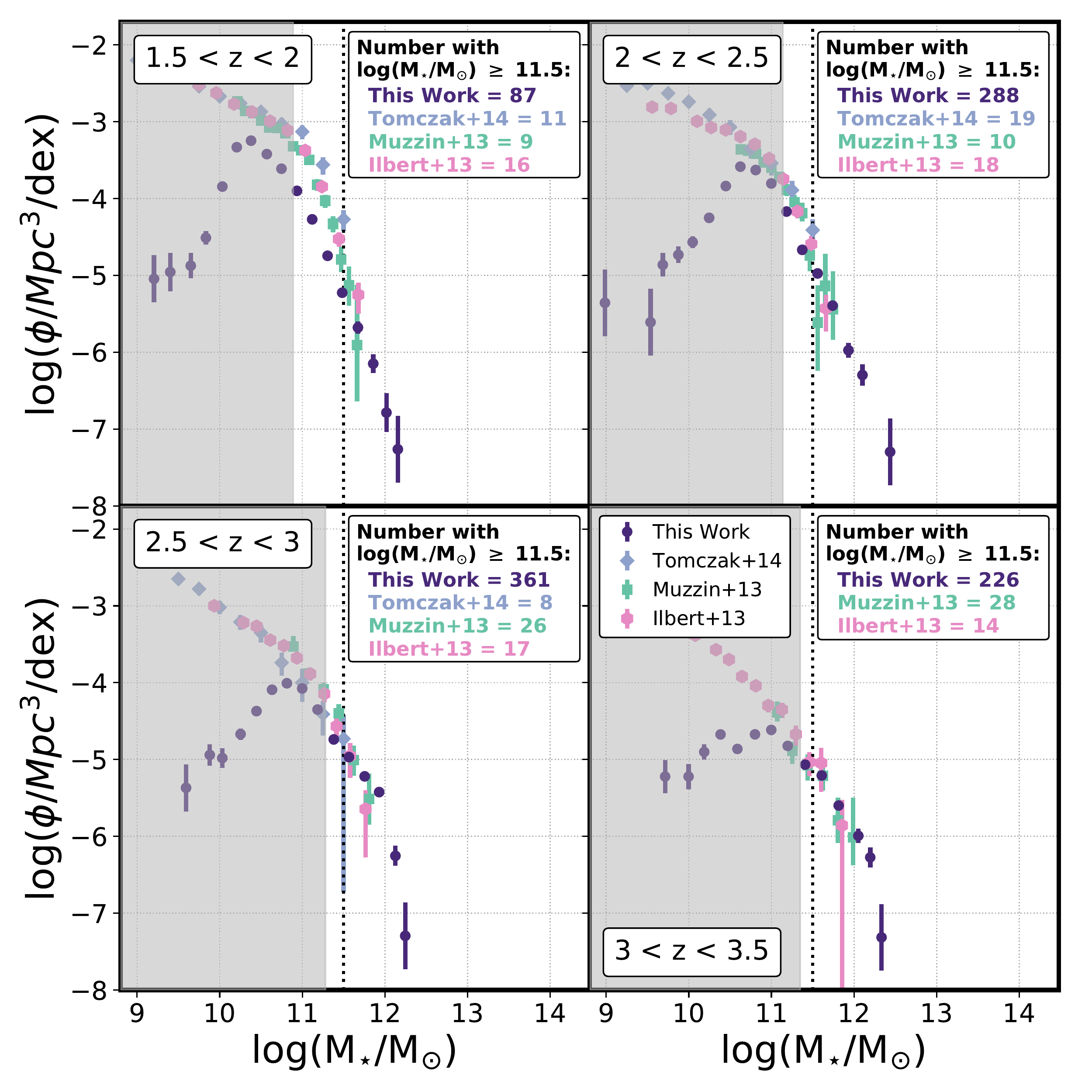} 
\caption{Galaxy stellar mass function for $1.5 < z < 3.5$ star-forming galaxies in the SHELA footprint. Error bars represent Poisson errors. The gray shaded regions represent masses below the 80\% mass completeness limit for each redshift bin. The vertical dotted line indicates log($M_\star$/$M_\odot$) = 11.5, the mass above which we compare to previous studies from the literature, restricted to only their star-forming samples. \textit{Note that in our $2 < z < 3.5$ redshift bins the size of our sample of massive (log($M_\star$/$M_\odot$) $\ge 11.5$) star-forming galaxies (as shown in the inset in each panel) is a over a factor of 10 larger than the samples from previously published studies.} We find that our star-forming galaxy stellar mass function shows fair agreement (within a factor of $\sim$ 2 to 4) with those from the literature.}
\label{smf_v_lit}
\end{center}
\end{figure*}

Over the past decade, three papers have risen to stand as the primary literature comparisons for intermediate redshift galaxy stellar mass functions. These are \cite{Tomczak2014} probing galaxies from $0.2 < z < 3$ and \cite{Muzzin2013} and \cite{Ilbert2013} which both explore the galaxy stellar mass function from $0 < z < 4$. In contrast to the SHELA survey, these works utilize small-area, deep surveys. The three works and their relation to the results in the SHELA footprint are described below and shown in Figure \ref{smf_v_lit}.

\cite{Tomczak2014} utilizes data from the FourStar Galaxy Evolution Survey (ZFOURGE; \citealt{Straatman2016}) with additional data from the Cosmic Assembly near-IR Deep Extragalactic Legacy Survey (CANDELS; \citealt{Grogin2011}, \citealt{Koekemoer2011}). These data span 316 arcmin$^2$, a significantly smaller area than the SHELA footprint, but are significantly deeper reaching an H$_{160}$ limiting magnitude of 26.5 mag. The primary leverage of a survey of this type is to investigate the low-mass end of the galaxy stellar mass function. To increase the area surveyed, the authors add data from the NEWFIRM Medium Band Survey (NMBS; \citealt{Whitaker2011}) spanning 1300 arcmin$^2$. Without the addition of larger area data, it would have been impossible for \cite{Tomczak2014} to place any constraints on the high-mass end of the stellar mass function. 

The larger 1.62 deg$^2$ UltraVista survey used by \cite{Muzzin2013} and \cite{Ilbert2013} is better equipped to study the high-mass end of the galaxy stellar mass function. While both studies use UltraVista imaging, their catalog construction, source selection, and SED fitting procedure differ. Both studies perform SED fitting with $\sim$30 filters in order to obtain photometric redshifts and galaxy stellar masses. Similarly, \cite{Muzzin2013} and \cite{Ilbert2013} investigate the evolution of the stellar mass function with redshift and find that the quiescent population evolves much more rapidly than the star-forming galaxy population. 

While each of these surveys places constraints on the high-mass end of the stellar mass function, cosmic variance remains high. From \cite{Moster2011}, the cosmic variance for $M_\star$ $>$ $10^{11}$$M_\odot$ galaxies in the \cite{Tomczak2014} study may be as much as 50-70\% from $1.5 < z < 3.5$, while \cite{Muzzin2013} and \cite{Ilbert2013} report cosmic variance to be 10-30\% for massive galaxies ($M_\star$ $>$ $10^{11}$$M_\odot$) at these redshifts. 

We compare the high-mass end (above our 80\% mass completeness limit) of the SHELA footprint star-forming galaxy stellar mass function to results from the literature (we scale the results from \cite{Muzzin2013} by a factor of 0.039 dex in stellar mass to convert from a Kroupa to Chabrier IMF) and find good agreement at all redshifts. Our results indicate that a significant population of massive star-forming galaxies is already in place at $z = 3$, where we find $\sim$39 galaxies per square degree ($\sim$6.28$\times$ $10^{-6}$ per Mpc$^3$) with $M_\star$ $>$ $10^{11}$$M_\odot$.

Our own study and those we compare with rely on SED fitting techniques to extract photometric redshifts and galaxy stellar masses, the results of which generally rely on the assumptions that are made. Agreement (disagreement) between results would then suggest a similarity (difference) in assumptions influencing the interpretation of the data. A detailed comparison of SED fitting codes and assumptions by \cite{Mobasher2015} showed that free parameters (such as redshift and star-formation history) can contribute a scatter in best-fit stellar mass of 0.136 dex across codes, while differences in model SED templates and the inclusion (or exclusion) of nebular emission lines can contribute a scatter of 0.2 dex and 0.3 dex, respectively. Our study utilizes FSPS templates with bursty SFHs for red galaxies and rising SFHs for blue galaxies. Additionally, our models implement nebular emission lines.  \cite{Tomczak2014} and \cite{Muzzin2013} both employ EAZY to obtain photometric redshifts and FAST \citep{Kriek2009} to fit for stellar masses using \cite{BruzualCharlot2003} models without nebular emission lines and exponentially declining SFHs. \cite{Ilbert2013} implement Le Phare (\citealt{Arnouts2002}, \citealt{Ilbert2006}) to fit for both photometric redshifts and stellar masses with \cite{BruzualCharlot2003} templates that include nebular emission lines, and exponentially declining SFHs. These differences in SED fitting methods lead to scatter in estimated stellar masses, which can lead to differences in the resulting galaxy stellar mass functions, particularly at the very steep high-mass end. As a result, it is challenging to disentangle discrepancies from fitting methods and differences in underlying physical processes contained in the implemented SED models when comparing observed stellar mass functions. An advantage of our large sample is that on average, we are more likely to uncover the true distribution of stellar masses of our galaxy population, while scatter increases with smaller sample sizes. 

\subsection{Comparing SHELA Footprint SMF to Theoretical Results}
\label{sec:smf_v_theory}
\subsubsection{Hydrodynamical Simulations}
In recent years, the Illustris Project (\citealt{Vogelsberger2014a}, \citealt{Vogelsberger2014b}, \citealt{Genel2014}) has emerged as a heavily relied upon hydrodynamical model tracing galaxy formation and evolution across cosmic time and on cosmological scales. The original Illustris Project utilized a $\sim$100$^3$Mpc$^3$ box with high resolution and reproduced observables such as the galaxy stellar mass function and cosmic star formation rate density. While this box size and model was the state-of-the-art for its time, the volume was simply too small to adequately study the mass assembly and properties of the most massive galaxies in the Universe. 

IllustrisTNG (\citealt{Pillepich2018b}, \citealt{Springel2018}, \citealt{Nelson2018}, \citealt{Naiman2018}, \citealt{Marinacci2018}), the newest rendition of the Illustris simulation, now includes a larger box size (TNG300 has $\sim$300$^3$Mpc$^3$) with updated physics. Such a large volume allows for the study of the most massive halos and the massive galaxies that live within them. Updated prescriptions for AGN feedback \citep{Weinberger2017} generate significantly more realistic massive galaxies with reasonable star formation rates, quenched fractions \citep{Donnari2019}, and gas fractions (\citealt{Weinberger2017}, \citealt{Pillepich2018a}). Without proper implementations of AGN feedback \citep{Weinberger2017}, the number density of massive galaxies \citep{Pillepich2018b} and their color distribution \citep{Nelson2018} do not match observations. If massive galaxies grow too large, without having star formation regulated by AGN feedback, this would artificially increase the number density of these systems. The solution, however, is not simply to increase the strength of AGN feedback to suppress star-formation as this would decrease the gas fraction in rich environments which was found to be low in the original Illustris simulation \citep{Weinberger2017}. Therefore, modified prescriptions, such as the kinetic-mode feedback implemented in IllustrisTNG are necessary to suppress star-formation while preserving the gas fraction and matching other observational constraints. 

When comparing the results from IllustrisTNG to that of the SHELA footprint (Figure \ref{smf_v_theory}), we utilize stellar mass functions from the largest volume of the IllustrisTNG series, TNG300, following \cite{Pillepich2018b}. Here, we adopt IllustrisTNG stellar masses measured within twice the half mass radius ($2\times r_{1/2}$) to ensure a one-to-one comparison between the way in which SHELA galaxy stellar masses are measured and how masses are extracted from the IllustrisTNG simulation. Additionally, we use the rescaled results of the simulation, denoted rTNG300, which is corrected for the reduced resolution of the $\sim$300$^3$Mpc$^3$ run compared to that of the $\sim$100$^3$Mpc$^3$ run \citep{Pillepich2018b}. To further ensure a fair comparison with the SHELA footprint results, we convolve the results from IllustrisTNG with the average SHELA footprint stellar mass error in each redshift bin ($\sim$0.16 dex), where this stellar mass error comes from our SED fitting procedure as described in Section \ref{sec:stellar_mass}. Finally, we adjust the total stellar mass function by the fraction of star-forming galaxies in each mass bin to obtain a star-forming galaxy stellar mass function. Quiescent galaxies are defined as those falling more than 1 dex below the galaxy main sequence, and those that don't meet that criterion are designated as star-forming \citep{Donnari2019}. 

At $1.5 < z < 2$ and $3 < z < 3.5$, we find that the rTNG300 model is in rough agreement (within a factor of 2 to 3) with that found in the SHELA footprint. In the $2 < z < 3$ bins the rTNG300 model agrees within a factor of 2 to 3 in the mass range $10^{11}$ - $5\times10^{11}$, but the discrepancy rapidly increases to a factor of $\sim$10 at the highest masses. It is important to note that although the increase in simulation volume to $\sim$300$^3$Mpc$^3$ is a significant milestone, this volume is only a small fraction of that of the SHELA footprint. Because of this, the results from rTNG300 do not probe the highest masses that we find in our massive galaxy sample. As computational resources continue to improve in coming years, large hydrodynamical simulation volumes will prove important in exploring the extreme high-mass ($M_\star$ $\gtrsim$ $10^{12}$$M_\odot$) galaxy regime. 

It is important to note that discrepancies in number density estimates for a particular mass bin are not the only consideration when comparing the results from simulation and observation. At the smallest number densities probed by rTNG300 ($\sim$3$\times$$10^{-5}$ Mpc$^3$/dex), galaxies from the simulation in our three highest redshift bins ($2 < z < 3.5$) are less massive than those in the SHELA footprint by $\sim$0.3 - 0.4 dex. Therefore, discrepancies in stellar mass function results may also be due to the way in which mass is assigned in the simulation. 

\subsubsection{Abundance Matching}
Typically, abundance matching assumes a direct relation between the size of a halo in a dark-matter only simulation and the stellar mass of a galaxy that would live in that halo (e.g., \citealt{Conroy2009a}, \citealt{Behroozi2010}, \citealt{Moster2013}). This method provides a direct link between simulated halo merger trees and observables. To improve upon this method, \cite{Behroozi2019} use a Monte Carlo based technique to probe the full range of observed quantities, such as the galaxy stellar mass function and cosmic star formation rate density to populate dark matter halos with realistic galaxies. Halos are populated first by assigning a star formation rate to a given halo and computing a stellar mass based on the assembly history of that halo. 

\begin{figure*}
\begin{center}
\includegraphics[width=\textwidth]{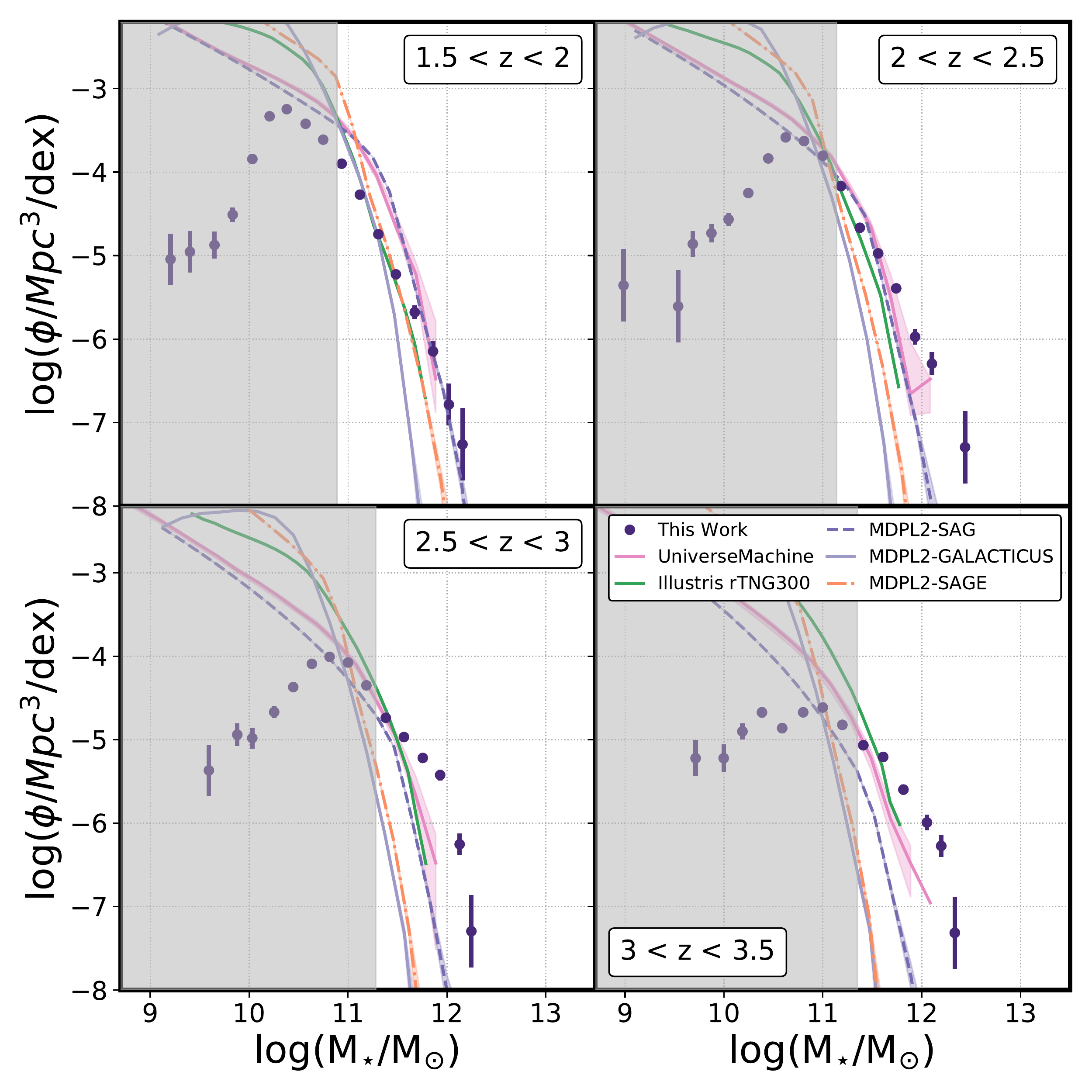} 
\caption{Comparison of the empirical galaxy stellar mass function for $1.5 < z < 3.5$ star-forming galaxies in the SHELA footprint to the stellar mass functions from three classes of theoretical models: hydrodynamical models from IllustrisTNG using rTNG300 \citep{Pillepich2018b}, abundance matching models from the UniverseMachine \citep{Behroozi2019} populating dark matter halos from \textit{Bolshoi-Planck}, and three different semi-analytic models (SAMs), namely \textsc{SAG} (radio-mode AGN feedback, \citealt{Cora2018}), \textsc{GALACTICUS} (radio-mode AGN feedback, \citealt{Benson2012}), and \textsc{SAGE} (quasar-mode AGN feedback, \citealt{Croton2016}) applied to the \textit{MDPL2} dark matter simulation. In our lowest redshift bin ($1.5 < z < 2$) results from all three model classes are in rough agreement with results from the SHELA footprint. In the $2 < z < 3$ bins, results from IllustrisTNG hydrodynamical models, UniverseMachine abundance matching, and the SAG semi-analytic model are within a factor of $\sim$2 - 10 to those from the SHELA footprint.  In this redshift bin, SAMs SAGE and GALACTICUS strongly underestimate the number density of massive galaxies by up to a factor of $\sim$100. In our highest redshift bin ($3 < z < 3.5$), the results from IllustrisTNG agree with those from the SHELA footprint within a factor of 2 - 3 and those from the UniverseMachine agree within a factor of 3 - 5. In this redshift bin all three SAMs underestimate the number density of massive galaxies by a factor of 10 - 1,000. We refer the reader to Section \ref{sec:smf_v_theory} for details. Error bars on SHELA footprint stellar mass function points represent Poisson errors, and the gray shaded region represents masses below the 80\% mass completeness limit for our SHELA footprint star-forming galaxy sample. All results from simulations have had mass and volume measures computed using (or scaled to) $h=0.7$, however no other aspects of the models have been scaled. The results from IllustrisTNG and the three SAMs have been convolved by the average stellar mass error for SHELA footprint galaxies in each redshift bin.}
\label{smf_v_theory}
\end{center}
\end{figure*}  

This matching method, the UniverseMachine \citep{Behroozi2019}, populates halos from the \textit{Bolshoi-Planck} and MultiDark-Planck2 (\textit{MDPL2}) dark matter-only simulations (\citealt{Klypin2016}, \citealt{RodriguezPuebla2016}) with galaxies from $0 < z < 10$. \textit{Bolshoi-Planck} and \textit{MDPL2} simulate 0.25$h^{-1}$Gpc and 1.0 $h^{-1}$Gpc, respectively, on each side \citep{Klypin2016}. While the box size of \textit{MDPL2} is larger than that of \textit{Bolshoi-Planck}, the mass resolution of \textit{MDPL2} ($1.5\times10^{9}M_{\odot}/h$) is poorer than \textit{Bolshoi-Planck} ($1.6\times10^{8}M_{\odot}/h$). The large box size of \textit{MDPL2} is used by \cite{Behroozi2019} primarily for probing the space of observed galaxy correlation functions, while \textit{Bolshoi-Planck}, with higher resolution, is better suited to this refined abundance matching method. The resulting galaxy stellar mass functions produced by the UniverseMachine are publicly available\footnote{\url{https://bitbucket.org/pbehroozi/universemachine}} for $0 < z < 10$. 

In Figure \ref{smf_v_theory} we compare the empirical SHELA footprint star-forming galaxy stellar mass function with the UniverseMachine stellar mass function. We correct the UniverseMachine stellar mass function, which contains all galaxies, by a factor of (1 - Quiescent Fraction) in each mass bin to obtain a star-forming galaxy stellar mass function. The quiescent fraction as a function of redshift and mass bin is given in the UniverseMachine data release. We do not apply any correction for systematic stellar mass error as this is already accounted for in the UniverseMachine data release. We find that in our lowest redshift bins ($1.5 < z < 2.5$) the UniverseMachine predicts number densities of massive galaxies within a factor of 2 to that found in the SHELA footprint. In the $2.5 < z < 3.5$ bins, the UniverseMachine under-predicts the number density of massive galaxies by a factor of 3 to 5.

As the method used by the UniverseMachine is heavily reliant on the available observations, this may be one explanation for this discrepancy. Additionally, discrepancies could point towards the relationship used by the UniverseMachine to relate baryonic properties to dark matter halos. Because the UniverseMachine calibrates results against small-area surveys, the strength of this abundance matching method is not at the high-mass end. With our study of massive galaxies across diverse environments at high redshifts, the results from the SHELA footprint may prove useful in calibrating abundance matching methods for massive galaxies. 

\subsubsection{Semi-Analytic Models}
\cite{Knebe2018} also utilize the \textit{MDPL2} dark matter simulation, and model the physics of baryons within dark matter halos by implementing semi-analytic models (SAMs). SAMs are models of galaxy evolution that track overall properties such as gas temperature and feedback processes in order to follow the growth of galaxies across cosmic time \citep{Somerville2015}. In contrast to more detailed numerical simulations, such as hydrodynamic simulations, SAMs are much less computationally expensive and can relatively easily be scaled to large volumes. 

The flexible and inexpensive nature of SAMs allows for different physical models of galaxy evolution and prescriptions for their growth over time to be explored. With this in mind, \cite{Knebe2018} applied three SAMs to the dark matter halos and merger trees from \textit{MDPL2}. The three SAMs, \textsc{GALACTICUS} \citep{Benson2012}, \textsc{SAG} \citep{Cora2018}, and \textsc{SAGE} \citep{Croton2016} differ in their prescriptions for cooling, star-formation, and AGN feedback, among others. These differences are described in detail in \cite{Knebe2018}, where they ultimately find that these differences lead to discrepancies between the SAMs in the number of massive galaxies produced by $z = 0.1$. \textsc{GALACTICUS} and \textsc{SAG} are found by \cite{Knebe2018} to over-predict the number of massive galaxies at $z = 0.1$ compared to \textsc{SAGE}, which produced a galaxy stellar mass function in agreement with results from SDSS at this redshift. They attribute these discrepancies to the differing treatment of AGN feedback among the models, which is implemented as radio-mode feedback in \textsc{GALACTICUS} and \textsc{SAG} and quasar-mode wind in \textsc{SAGE}. They also note that the differences seen in the stellar mass function may be attributed to different prescriptions for galaxy mergers and interactions, which become increasingly important in the rich environments where many massive galaxies reside, as well as differing prescriptions for the treatment of satellite galaxies. 

Results from applying the \textsc{GALACTICUS}, \textsc{SAGE}, and \textsc{SAG} SAMs to the \textit{MDPL2} dark matter halos are publicly available\footnote{\url{http://www.skiesanduniverses.org}}, and we utilize results for the z $\sim$ 2, 2.5, 3, and 3.5 snapshots to compare with the results from galaxies in the SHELA footprint (Figure \ref{smf_v_theory}). Each galaxy in a snapshot has separate stellar masses and star formation rates published for the disk and spheroid components of the galaxy. We combine these two values to obtain a single stellar mass and star formation rate value for each galaxy. We further convolve this mass by the average stellar mass error of SHELA footprint galaxies (determined from our SED fitting procedure, see Section \ref{sec:stellar_mass}) in each redshift bin ($\sim$ 0.16 dex) to provide a more realistic comparison with our empirical result  \citep{Kitzbichler2007}. We compute the specific star formation rate for each object and keep only star-forming galaxies having sSFR > $10^{-11}~yr^{-1}$, consistent with our selection of star-forming galaxies in the SHELA footprint. Stellar masses along with the box volume at each snapshot are used to compute the star-forming galaxy stellar mass function for each SAM following the method of \cite{Tomczak2014} where every galaxy in a particular redshift bin has the same comoving volume term.

We find that in the $1.5 < z < 3.5$ bin, all three SAMs are in good agreement with the results from the SHELA footprint. In the $2 < z < 3$ redshift bins, SAG agrees with the results from the SHELA footprint to within a factor of 10, however at these redshifts SAGE and GALACTICUS under-estimate the number density of star-forming galaxies by a factor of 10 - 100. In the highest redshift bin ($3 < z < 3.5$), the three SAMs severely under-estimate the number density of massive star-forming galaxies by up to a factor of 1,000, with SAG showing less discrepancy than SAGE and GALACTICUS. This result is in agreement with \cite{Asquith2018}, who find that \textsc{SAGE} and \textsc{SAG} underestimate the number density of massive galaxies when compared to intermediate redshift observations. \cite{Asquith2018} perform their study using a different underlying dark matter model than that of \cite{Knebe2018}, which points to the discrepancy lying with the SAMs themselves, rather than the dark-matter simulation. Similar discrepancies between SAM results and observations were found by \cite{Conselice2007}, who showed a factor of 100 difference between massive galaxy number densities from the Millennium Run SAM \citep{DeLucia2006} and observations in the Palomar/DEEP2 survey (e.g., \citealt{Davis2003}, \citealt{Giavalisco2004}, \citealt{Bundy2005}, \citealt{Davis2007}) at $z\sim2$. 

The discrepancies found between stellar mass functions from the SHELA footprint and those from the SAMs are not exclusively linked to the models under-predicting the number density of massive galaxies. An alternative is that the SAMs simply do not produce galaxies that are massive enough (they are under-massive by up to $\sim$1.5 dex at the lowest number densities) pointing to the implementation of baryonic physics as an underlying cause. We find that at $1.5 < z < 3.5$, the results from GALACTICUS and SAGE are in agreement despite their implementation of quasar-mode and radio-mode AGN feedback respectively. SAG, which utilizes quasar-mode feedback, produces more massive galaxies in all redshift bins than both GALACTICUS and SAGE. This suggests that different implementations of baryonic physics, such as prescriptions for satellite treatment, cooling, and details of the particular AGN feedback mode may play an important role.

\section{Summary}
\label{sec:summary}
We present a comprehensive study of the high-mass end of the star-forming galaxy stellar mass function at cosmic noon ($1.5 < z < 3.5$), an important epoch in the growth of galaxies, and their constituent stars and black holes. The main strengths of this study are that the sample is drawn from a very large area (17.2 deg$^2$, corresponding to a colossal comoving volume of $\sim$0.44 Gpc$^3$ from $1.5 < z < 3.5$) thereby limiting cosmic variance and encompassing a wide range of environments (fields, groups, and proto-clusters). The resulting sample of massive star-forming galaxies (5,352 galaxies with log($M_\star$/$M_\odot$) $>$ 11, of which 962 have log($M_\star$/$M_\odot$) $>$ 11.5) at $1.5 < z < 3.5$ in this large volume is a factor of 10 larger than previous studies, thereby providing robust statistics in the high-mass galaxy regime. To further increase the strength of our results, we also test uncertainties in SED fitting and characterize how well galaxy parameters can be recovered. Our results are summarized below. 

\begin{itemize}
\item  \noindent We find that the star-forming galaxy stellar mass function in the SHELA footprint from $1.5 < z < 3.5$ is steeply declining at the high-mass end and probes values as high as $\sim$$10^{-4}$ Mpc$^3$/dex and as low as $\sim$5$\times$$10^{-8}$ Mpc$^3$/dex, across a stellar mass range of log($M_\star$/$M_\odot$) $\sim$ 11 - 12. With our statistically significant sample of high-mass star-forming galaxies and improved SED fitting procedure, we place some of the strongest constraints to date on the high-mass end of the star-forming galaxy stellar mass function from $1.5 < z < 3.5$.
\\
\item \noindent We compare our results to previous observational studies from \cite{Tomczak2014}, \cite{Muzzin2013}, and \cite{Ilbert2013}. Our sample of massive star-forming galaxies (962 galaxies with log($M_\star$/$M_\odot$) $>$ 11.5) is more than a factor of 10 larger than the samples presented in these previous studies. Our results are therefore more statistically robust and suffer less from uncertainties due to cosmic variance. We find that our results are in good agreement with those from the literature in all of our redshift bins. 
\\
\item \noindent We compare our results to different numerical models of galaxy evolution, including hydrodynamical simulations from IllustrisTNG (\citealt{Pillepich2018b}, \citealt{Springel2018}, \citealt{Nelson2018}, \citealt{Naiman2018}, \citealt{Marinacci2018}), abundance matching from the UniverseMachine \citep{Behroozi2019}, and three SAMs (SAG \citep{Cora2018}, SAGE  \citep{Croton2016}, and GALACTICUS \citep{Benson2012}). In our lowest redshift bin ($1.5 < z < 2$) we find that all models are in rough agreement with the results from the SHELA footprint. For $2 < z < 3$, we find that the IllustrisTNG hydrodynamical models, UniverseMachine abundance matching models, and SAG semi-analytic model results are within a factor of $\sim$ 2 to 10 of the SHELA footprint results at the high-mass end of the galaxy stellar mass function. In the highest redshift bin ($3 < z < 3.5$), results from IllustrisTNG and the UniverseMachine are within a factor of 2 to 5 to those from the SHELA footprint. At $2 < z < 3.5$, SAMs SAGE and GALACTICUS severely underestimate the number density of galaxies by a factor of 10 - 1,000, and in the highest redshift bin ($3 < z < 3.5$), a factor of 100 discrepancy is also seen with SAG. These large discrepancies highlight the challenges that SAMs face in implementing baryonic physics that can reproduce observed galaxy relations simultaneously at low- ($z = 0$), and high-redshifts ($z = 2 - 4$).

\end{itemize}

\vspace{5mm}
S.S., S.J, and J.F. gratefully acknowledge support from the University of Texas at Austin, as well as NSF grants AST 1614798 and 1413652. S.S., S.J, J.F., M.S., and S.F. acknowledge generous support from The University of Texas at Austin McDonald Observatory and Department of Astronomy Board of Visitors. The authors wish to thank Christopher Conselice for his constructive comments, Annalisa Pillepich and Mark Vogelsberger for providing IllustrisTNG SMF results and useful comments, Peter Behroozi for helpful comments, and Andrew Benson, Sofia Cora, and Darren Croton for their feedback regarding comparisons with semi-analytic models. L.K. and C.P. acknowledge support from the National Science Foundation through grant AST 1614668. L.K. thanks the LSSTC Data Science Fellowship Program; her time as a Fellow has benefited this work. The Institute for Gravitation and the Cosmos is supported by the Eberly College of Science and the Office of the Senior Vice President for Research at the Pennsylvania State University. This publication uses data generated via the Zooniverse.org platform, development of which is funded by generous support, including a Global Impact Award from Google, and by a grant from the Alfred P. Sloan Foundation.

\bibliographystyle{mnras}

\bsp	
\label{lastpage}
\end{document}